# Background Conditions Influence the Estimated Cloud Radiative Effects of Anthropogenic Aerosol Emissions from Different Source Regions

**Benjamin S. Grandey[1], and Chien Wang[2,1]**

[1]Center for Environmental Sensing and Modeling, Singapore-MIT Alliance for Research and Technology, Singapore, Singapore.

[2]Center for Global Change Science, Massachusetts Institute of Technology, Cambridge, Massachusetts, USA.

Corresponding authors: Benjamin S. Grandey (benjamin@smart.mit.edu) and Chien Wang (wangc@mit.edu)

**Key Points:**

- The responses of the shortwave and the longwave cloud radiative effects to aerosol emissions are sub-linear.
- The cloud radiative effects of anthropogenic aerosol emissions from a given source region are sensitive to emissions from other regions.
- The cloud radiative effects are weaker under present-day background conditions compared with under natural background conditions.

**Abstract**

Using the Community Earth System Model, with the Community Atmosphere Model version 5.3, we investigate the cloud radiative effects of anthropogenic aerosols emitted from different source regions and global shipping. We also analyse aerosol burdens, cloud condensation nuclei concentration, liquid water path, and ice water path. Due to transboundary transport and sub-linearity in the response of clouds to aerosols, the cloud radiative effects of emissions from a given source region are influenced by emissions from other source regions. For example, the shortwave cloud radiative effect of shipping is $-0.39 \pm 0.03$ W m$^{-2}$ when other anthropogenic emissions sources are present (the "present-day background" assumption) compared with $-0.60 \pm 0.03$ W m$^{-2}$ when other anthropogenic emissions sources are absent (the "natural background" assumption). In general, the cloud radiative effects are weaker if present-day background conditions are assumed compared with if natural background conditions are assumed. Assumptions about background conditions should be carefully considered when investigating the climate impacts of aerosol emissions from a given source region.

**Plain Language Summary**

Particulate air pollutants interact with clouds, influencing Earth's radiation budget. We use a global climate model to investigate the cloud radiative effects of particulate emissions from different source regions and global shipping. Due to non-linearity, the radiative effects of emissions from one region are influenced by emissions from other regions – this should be taken into account when investigating the climate impacts of particulate pollutants.

**1 Introduction**

By acting as cloud condensation nuclei, aerosol particles play an important role in cloud microphysics. Therefore, perturbing the availability or properties of aerosol particles may influence cloud properties and Earth's radiation budget. Assuming constant liquid water content within a cloud, Twomey (1974, 1977) proposed that increasing the availability of cloud condensation nuclei would lead to a larger number of smaller cloud droplets, increasing the albedo of the cloud, causing more shortwave radiation to be reflected to space: this is referred to as the cloud albedo effect, also known as the first indirect effect. Albrecht (1989) suggested that the decrease in droplet size would suppress precipitation, increasing the lifetime of the cloud, potentially increasing cloud cover, again causing more shortwave radiation to be reflected to space: this is referred to as the cloud lifetime effect, also known as the second indirect effect. Subsequently, other indirect effects have been proposed, as reviewed by Lohmann and Feichter (2005), Tao et al. (2012), Rosenfeld et al. (2014), and Fan et al. (2016). However, despite extensive research, aerosol effects on clouds and precipitation remain poorly understood (Stevens & Feingold, 2009).

Cloud radiative effects play a critical role in the net effective radiative forcing of anthropogenic aerosols. Uncertainty in estimates of the cloud radiative effects causes uncertainty in estimates of aerosol effective radiative forcing (Quaas et al., 2009; Shindell et al., 2013), contributing to substantial uncertainty in estimates of climate sensitivity and projections of future climate change (Andreae et al., 2005; Kiehl, 2007). Regional heterogeneity in the cloud radiative effects also influences both climate sensitivity (Shindell, 2014) and precipitation distribution (Wang, 2015). It is therefore important to improve understanding of the strength and regional distribution of the cloud radiative effects.

One challenge is that the cloud radiative effects of aerosols are likely non-linear. Due to this non-linearity, the cloud radiative effects are sensitive to the background conditions that exist prior to a perturbation in aerosol emissions. This non-linearity has implications for constraining the strength of the cloud radiative effects of anthropogenic aerosol emissions: aerosol-cloud relationships under present-day conditions may not be suitable for quantifying the sensitivity of clouds to anthropogenic aerosol emissions (Ghan et al., 2016; Penner et al., 2011; see also Gryspeerdt et al., 2017); and uncertainty in pre-industrial aerosol emissions may contribute a large source of uncertainty to the cloud radiative effects of anthropogenic aerosol emissions (Carslaw et al., 2013; Regayre et al., 2018; Wilcox et al., 2015). The non-linearity also has implications for the cloud radiative effects of other sources of aerosols: for example, ignoring the interannual variability of aerosol emissions from wildfires may lead to a systematic overestimation of the cloud radiative effects (Grandey et al., 2016b).

With tropospheric lifetimes of days to weeks, aerosols emitted from a given source region are often transported to other receptor regions. Therefore, in addition to influencing clouds locally, aerosol emissions from a given source region may influence clouds over receptor regions. Furthermore, by influencing background conditions, aerosol emissions from a given source region may also modulate the influence of aerosol emissions from other source regions, and vice versa. This interplay – between regional emissions, transboundary transport, background conditions, and non-linearity of the cloud radiative effects – is the focus of our study.

Using four different models (three of which included cloud radiative effects), Bellouin et al. (2016) investigated the specific radiative forcing – defined as "the radiative forcing per unit change in mass emitted" (Bellouin et al., 2016) – associated with anthropogenic aerosol emissions from Europe, East Asia, and shipping. They found the specific radiative forcing to be dependent on emission location: in comparison to the emissions from Europe and shipping, the emissions from East Asia exert a weaker specific radiative forcing, likely due to more polluted background conditions over East Asia.

Chen et al. (2018) investigated the radiative effects of anthropogenic organic carbon and black carbon aerosol emissions from Asia and North America, using a modified version of the Community Earth System Model (CESM). They identified non-linearity in the cloud radiative effect, especially near strong emissions sources of organic carbon aerosol.

Yang et al. (2017) investigated the impacts of sulfur emissions from different source regions, using CESM. By tagging the sulfur emitted from the different source regions, they identified the contributions to near-surface sulfate aerosol concentration and the direct radiative effect. Using an additional simulation, they quantified the net cloud radiative effect associated with a 20% reduction in global sulfur emissions. They subsequently decomposed this cloud radiative effect into contributions from different source regions, by scaling the cloud radiative effect "in a grid column by regional source contributions to sulfate mass concentration reduction averaged from the surface layer to 850 hPa", providing "a computationally efficient method of quantifying regional [indirect radiative forcing] that has a higher signal to noise as compared to regional perturbation simulations" (Yang et al., 2017). In general agreement with Bellouin et al. (2016), Yang et al. (2017) found that aerosol emissions from more polluted regions in the Northern Hemisphere exert a weaker specific radiative forcing than emissions from cleaner regions in the Southern Hemisphere.

Although we also use CESM, our methodology differs from that of Yang et al. (2017). One substantial difference is that we perturb regional emissions in different simulations, calculating the cloud radiative effects using a "radiative flux perturbation" approach (Haywood et al., 2009). This allows us to explore non-linearity in the cloud radiative effects. A second substantial difference is that we investigate the influence of background conditions.

Previous studies have investigated the cloud radiative effects of aerosol emissions from different source regions (Bellouin et al., 2016; Chen et al., 2018; Yang et al., 2017). Other studies have investigated the influence of background conditions, especially uncertainty relating to pre-industrial aerosol emissions (Carslaw et al., 2013; Regayre et al., 2018; Wilcox et al., 2015). However, to our knowledge, no previously published study has used a global climate model to explore the interplay between regional emissions and background conditions.

Our primary scientific question is as follows: How do assumptions about background conditions influence the cloud radiative effects of aerosol emissions from different source regions?

## 2 Methodology

### 2.1 Model configuration

We use the Community Earth System Model (CESM) version 1.2.2 and the Community Atmosphere Model version 5.3 (CAM5.3), with CAM5.3's default three-mode aerosol module (MAM3; Liu et al., 2012), aerosol activation scheme (Abdul-Razzak & Ghan, 2000), and stratiform cloud microphysics scheme (Gettelman et al., 2010; Morrison & Gettelman, 2008). CAM5.3 is run at a horizontal resolution of $1.9°\times2.5°$ with 30 levels in the vertical direction. The base configuration for all simulations follows the "F_2000_CAM5" component set, with prescribed greenhouse gas concentrations and prescribed sea-surface temperatures (SSTs) following year-2000 climatological values. The simulations differ only in their aerosol (including aerosol precursor) emissions, following the scenarios described in Section 2.2 below.

By comparing pairs of prescribed-SST simulations with different aerosol emissions, the radiative effects of aerosols can be diagnosed using the "radiative flux perturbation" approach (Haywood et al., 2009). A diagnostic "clean-sky" radiation call facilitates decomposition of the direct radiative effect, clean-sky shortwave cloud radiative effect, longwave cloud radiative effect, and surface albedo radiative effect (Ghan, 2013). In this study, we analyse the two cloud radiative effect components.

Each simulation includes a two-year spin-up period that is discarded. The analysis period for the *All1* and *All0* simulations (see below) is 60 simulation years; the analysis period for the other simulations is 30 simulation years.

### 2.2 Aerosol emissions scenarios

In order to investigate the science question identified above, we analyse results from 22 different simulations following 22 different emissions scenarios. These scenarios can be considered as two individual reference scenarios plus two additional groups, each containing ten scenarios:

1. *"All1"* uses year-2000 aerosol and aerosol precursor emissions. The emitted mass concentrations and number concentrations follow the year-2000 emissions from the default historical (1850–2005) emissions files for MAM3 (Liu et al., 2012). The anthropogenic and wildfire emissions of sulfur dioxide, organic carbon aerosol, and black carbon aerosol are based on the historical emissions of Lamarque et al. (2010). Emissions of dimethyl sulfide and continuous volcanic emissions of sulfur dioxide follow Dentener et al. (2006). Approximately 2.5% of the sulfur dioxide is emitted as primary sulfate. (Further discussion of the primary sulfate emissions can be found in Text S1 of the Supporting Information.) Information about injection height profiles, the size distributions used to derive number emissions, and emissions of volatile organic compounds can be found in the Supplement of Liu et al. (2012).

2. *"All0"* has no anthropogenic emissions of sulfur dioxide, primary sulfate aerosol, black carbon aerosol, or organic carbon aerosol. Wildfire emissions and natural emissions remain the same as in *All1*. Comparison of *All1* with *All0* reveals the total impact of anthropogenic aerosol emissions.

3. *"Ship1", "EAs1", "NAm1", "Eur1", "AfME1", "CAs1", "SAs1", "SEAs1", "SAm1",* and *"ANZ1"* are a group of ten scenarios with no anthropogenic emissions apart from either (a) global shipping emissions (*Ship1*) or (b) land-based anthropogenic emissions from a given source region (Fig. 1). For example, *EAs1* has land-based anthropogenic emissions from East Asia but has no other anthropogenic emissions. Apart from the targeted emissions, all other emissions remain the same as in *All0*. We refer to these ten scenarios as the *"Θ1"* group of scenarios, where *Θ* refers to a given source region (or shipping). Comparison of a *Θ1* scenario with *All0* reveals the impact of anthropogenic emissions from a given source region, assuming there are no other anthropogenic emissions. We refer to this assumption as the "natural background" assumption.

4. *"Ship0", "EAs0", "NAm0", "Eur0", "AfME0", "CAs0", "SAs0", "SEAs0", "SAm0",* and *"ANZ0"* are a group of ten scenarios with either (a) no shipping emissions globally (*Ship0*) or (b) no land-based anthropogenic emissions from a given source region (Fig. 1), while all other emissions remain at year-2000 levels. For example, *EAs0* has no land-based anthropogenic emissions from East Asia. Apart from the targeted emissions, all other emissions remain the same as in *All1*. We refer to these ten scenarios as the *"Θ0"* group of scenarios. Comparison of a *Θ0* scenario with *All1* reveals the impact of anthropogenic emissions from a given source region, assuming all other emissions remain at year-2000 levels. We refer to this assumption as the "present-day background" assumption.

The key distinction between the *Θ1* scenarios and the *Θ0* scenarios is that *Θ1-All0* reveals the impact of a subset of anthropogenic emissions under natural background conditions (i.e. all other anthropogenic emissions are "switched off"), while *All1-Θ0* reveals the impact of a subset of anthropogenic emissions under present-day background conditions (i.e. all other anthropogenic emissions remain "switched on"). This allows us to investigate non-linearity in the radiative effects of aerosol emissions. Due to such non-linearity, we would not necessarily expect the impacts revealed by *Θ1-All0* to necessarily be the same as those revealed by *All1-Θ0*.

Global emissions of sulfur dioxide, organic carbon aerosol, and black carbon aerosol for each scenario, relative to *All0* and *All1*, are shown in Figs. 2, S1, and S2.

# 3 Results

In order to inform discussion of the cloud radiative effects, we first analyse aerosol burdens and cloud condensation nuclei concentration. We also analyse liquid water path, which is negatively correlated with the shortwave cloud radiative effect, and ice water path, which is positively correlated with the longwave cloud radiative effect (Grandey et al., 2018a).

## 3.1 Aerosol burdens

Aerosol burdens – also known as "column loadings" – reveal the total amount of a given species of aerosol in an atmospheric column. We focus primarily on the sulfate aerosol burden, based on the assumption that anthropogenic sulfate dominates the anthropogenic cloud radiative effects, our primary interest.

The global-mean sulfate aerosol burden ($Burden_{SO4}$) has a natural baseline of 0.97 mg($SO_4$) m$^{-2}$ (Fig. 3), approximately 0.5 Tg($SO_4$) globally. Anthropogenic sulfur emissions contribute +2.00 mg($SO_4$) m$^{-2}$, twice the natural baseline. The anthropogenic component of the global-mean primary organic matter aerosol burden is +0.33 mg m$^{-2}$, 41% of the natural baseline (Fig. S3); the anthropogenic component of the global-mean black carbon aerosol burden is +0.089 mg m$^{-2}$, 130% of the natural baseline (Fig. S4).

For the burdens of all three aerosol species, the contributions from the different source regions combine linearly (Figs 3, S3, and S4): $\Sigma_\Theta(\Theta 1-All0)$, under the natural background assumption, is very similar to $\Sigma_\Theta(All1-\Theta 0)$, under the present-day background assumption; and the relative contributions from the different source regions appear to be independent of the assumption about background conditions.

The largest anthropogenic contributors to $Burden_{SO4}$ are North America, Africa and the Middle East, East Asia, and Europe (Fig. 3): taken together, these regions are responsible for approximately two-thirds of the anthropogenic component of $Burden_{SO4}$. For most source regions, the percentage contribution to the anthropogenic component of $Burden_{SO4}$ is generally comparable to the percentage contribution to anthropogenic sulfur dioxide emissions (Fig. 2). Exceptions include East Asia, which contributes 21 % to the anthropogenic sulfur dioxide emissions but only 14–15 % to the anthropogenic component of $Burden_{SO4}$, and shipping, which contributes 11 % to the anthropogenic emissions but only 7 % to the anthropogenic component of $Burden_{SO4}$. Conversely, Africa and the Middle East contributes disproportionately: despite contributing only 10 % to the anthropogenic emissions, Africa and the Middle East contributes 17–18 % to the anthropogenic component of $Burden_{SO4}$. These differences in the influence of regional emissions on $Burden_{SO4}$ are likely due to regional variation in the lifetime of sulfate aerosol, dependent on the cloud and precipitation distribution (Yang et al., 2017); other factors, such as regional variation in the conversion efficiency of sulfur dioxide to sulfate aerosol, may also play a role.

Transboundary transport is evident for all three aerosol species (Figs. 4, S5, and S6). For example, North America, Europe, Central Asia, East Asia, and South Asia all make substantial contributions (>5%) to total $Burden_{SO4}$ over Greenland (Fig. 4). Therefore, due to non-linearity in the response of clouds to aerosols, it is to be expected that the radiative effects of aerosols emitted from one region may be influenced by emissions from other regions.

### 3.2 Cloud condensation nuclei concentration

Aerosol particles have the potential to become activated as cloud condensation nuclei (CCN). In general, the more hydrophilic and the larger a given aerosol particle is, the more likely it is to be activated. Whether or not the particle becomes activated depends on the supersaturation, which in turn is influenced by competition for water vapour between the aerosol particles present (Rothenberg et al., 2018; Rothenberg & Wang, 2016, 2017). However, to facilitate analysis, the CCN concentration results analysed here are for a fixed supersaturation of 0.1 %.

We choose to analyze the CCN concentration in model level 24, corresponding to a pressure height of approximately 860 hPa, in the lower troposphere. This height should be representative of conditions at cloud base for liquid water clouds. Furthermore, we have previously found that the spatial pattern of CCN concentration in the lower troposphere is similar to that near the surface and in the mid-troposphere (Grandey et al., 2018a); therefore, the analysis should not depend strongly on the choice of specific model level.

The percentage contributions of each source region to the total CCN concentration (Fig. S7) are comparable to those of each source region to total $Burden_{SO4}$ (Fig. 4). For example, North America, Europe, Central Asia, and East Asia (but not South Asia) all make substantial contributions (>5%) to total CCN concentration over Greenland (Fig. S7). (However, in contrast to the total $Burden_{SO4}$ results, shipping also makes a substantial contribution to CCN concentration over Greenland.) Globally, for each source region, the percentage contribution to the anthropogenic component of CCN concentration (Fig. S8) is generally similar to the percentage contribution to the anthropogenic component of $Burden_{SO4}$ (Fig. 3). The close relationship between $Burden_{SO4}$ and CCN concentration is not surprising, because sulfate aerosol, which is hydrophilic, is a dominant source of CCN.

As was the case for the aerosol burdens, the absolute contributions of the different source regions to the CCN concentration do not depend strongly on the assumption about background conditions, indicating linearity (Fig. S8). Furthermore, $\Sigma_\Theta(\Theta 1-All0)$, under the natural background assumption, is very similar to $\Sigma_\Theta(All1-\Theta 0)$, under the present-day background assumption. The absolute differences in the zonal-mean CCN concentrations also do not depend strongly on background conditions, again indicating linearity (Fig. S9).

However, although the *absolute* differences in CCN concentration may be independent of background conditions (Fig. S9), the *relative percentage changes* are strongly dependent on background conditions (Fig. 5). For emissions from a given region, the percentage change in CCN concentration is much weaker under the present-day background assumption compared with under the natural background assumption. For example, at approximately 35°N, anthropogenic emissions from East Asia perturb the CCN concentration by approximately 50 % under the natural background assumption, twice as much as under the present-day background assumption (Fig 5a); shipping emissions perturb the CCN concentration by approximately 10 % under the natural background assumption, three times as much as under the present-day background assumption (Fig 5d).

The relative percentage changes in CCN concentration exhibit sub-linearity (Fig. 5). This sub-linearity will likely drive sub-linearity in liquid water path and the shortwave cloud radiative effect, both of which are discussed below.

### 3.3 Liquid water path

Globally, the anthropogenic aerosol emissions drive increases in grid-box liquid water path (Fig. 6). Under the natural background assumption, the largest increases are associated with emissions from East Asia, North America, shipping, Europe, and Africa and the Middle East. Under the present-day background assumption, the largest increases are associated with the same five regions, albeit in a different order.

For emissions from a given region, latitudes with the largest differences in zonal-mean liquid water path (Fig. 7) generally correspond to those with the largest percentage changes in CCN concentration (Fig. 5). For example, East Asian emissions exert the strongest influence at approximately 30°N, North American emissions at approximately 45°N, and European emissions at approximately 50°N.

Evidence of weak sub-linearity can be found in both the zonal-mean (Fig. 7) and the global (Fig. 6) liquid water path results. For example, anthropogenic emissions from Europe drive an increase of +0.61±0.05 g m$^{-2}$ under the natural background assumption, compared with +0.45±0.06 g m$^{-2}$ under the present-day background assumption. $\Sigma_\Theta(All1-\Theta 0)$ is 80±9 % of $\Sigma_\Theta(\Theta 1-All0)$. However, the $\Sigma_\Theta(All1-\Theta 0)-\Sigma_\Theta(\Theta 1-All0)$ difference is statistically insignificant.

### 3.4 Shortwave cloud radiative effect

Globally, the anthropogenic aerosol emissions exert a negative clean-sky shortwave cloud radiative effect ($\Delta CRE_{SW}$) of −2.63±0.02 W m$^{-2}$, cooling the climate system (Fig. 8). This cooling effect is stronger than the values of −2.0 to −2.1 W m$^{-2}$ reported by Ghan et al. (2012) and by Grandey et al. (2018a), because Ghan et al. and Grandey et al. quantified 2000–1850 differences whereas we quantify the impact of all anthropogenic aerosol emissions. (The value of −2.1 W m$^{-2}$ reported by Grandey et al. (2018a) corresponds to the same model version and a similar configuration to that used in this study.)

Under both the natural background assumption and the present-day background assumption, the five largest contributing sources, listed in order of strength, are shipping, East Asia, North America, Africa and the Middle East, and Europe – the five largest sulfur emissions sources, albeit in a different order. Shipping contributes disproportionately, contributing 11% to the anthropogenic sulfur emissions (Fig. 2) but 19–20% to combined anthropogenic $\Delta CRE_{SW}$ (Fig. 8). We suggest four possible reasons for the disproportionate impact of the shipping emissions, three of which are related to the fact the shipping emissions occur over ocean: first, the ocean surface is much darker than most land regions, so a cloud will likely exert a larger radiative effect over ocean compared with over land; second, clouds in clean remote ocean regions may often be more susceptible to aerosol perturbations than clouds in polluted continental regions are; third, marine stratocumulus clouds may be particularly susceptible to aerosol perturbations; fourth, the shipping emissions are spread out over a larger geographical area, potentially leading to a stronger radiative effect due to sub-linearity in the $\Delta CRE_{SW}$ response.

Sub-linearity is evident in the global-mean $\Delta CRE_{SW}$ results (Fig. 8). For example, shipping emissions produce a cooling effect of −0.60±0.03 W m$^{-2}$ under the natural background assumption compared with a weaker cooling effect of −0.39±0.03 W m$^{-2}$ under the present-day background assumption. Similarly, with the exception of Australia and New Zealand (which produces insignificant cooling), the cooling effect of all land-based source regions is weaker under the present-day background assumption compared with under the natural background assumption.

When the contributions from the different sources are summed, the $\Sigma_\Theta(All1-\Theta0)-\Sigma_\Theta(\Theta1-All0)$ difference is both large and statistically significant (+1.13±0.27 W m$^{-2}$). $\Sigma_\Theta(All1-\Theta0)$, under the present-day background assumption, is only 63±8 % of $\Sigma_\Theta(\Theta1-All0)$, under the natural background assumption.

Interestingly, the relative contribution of each emissions source to the summation does not depend strongly on the assumption about background conditions (Fig. 8). For example, shipping emissions contribute 19±2 % under the natural background assumption, which is very similar to 20±3 % under the present-day background assumption. When background conditions change, the *relative* importance of each source does not change substantially, although the *absolute* importance of each source changes.

Sub-linearity is also evident in the zonal-mean $\Delta CRE_{SW}$ results (Fig. 9). For example, the anthropogenic emissions from Europe lead to a peak cooling of approximately −2 W m$^{-2}$ under the natural background assumption compared with −1 W m$^{-2}$ under the present-day background assumption (near 45–60°N in Fig. 9c). For the other four of the five largest sources of sulfur dioxide emissions, the influence of background conditions is not as strong (Fig. 9a, b, d, and e). Nevertheless, the peak zonal-mean cooling is always weaker under the present-day background assumption compared with under the natural background assumption (e.g. near 30°N in Fig. 9a). However, at latitudes with weak zonal-mean cooling, the difference between the natural background and present-day background results is generally insignificant (e.g. in the Southern Hemisphere in Fig. 9a): with the exception of *All1–All0* (Fig. 9f), the zonal-mean results are often noisy.

3.5 Ice water path

Both sulfate and dust act as ice nuclei in the ice and mixed-phase cloud microphysics scheme (Gettelman et al., 2010; Liu et al., 2007). However, we lack diagnostics of ice nuclei availability. We proceed to analyse ice water path, a cloud property that is influenced strongly by ice nuclei availability.

Globally, the anthropogenic aerosol emissions drive an increase in ice water path (Fig. 10). The differences in zonal-mean ice water path are particularly noisy (Fig. S10). The influence of sub-linearity on the global-mean results is very strong: $\Sigma_\Theta(\Theta1-All0)$ is much larger than *All1-All0*, while $\Sigma_\Theta(All1-\Theta0)$ is much smaller and statistically insignificant (Fig. 10). We would expect the strong sub-linearity in the ice water path results to correspond to strong sub-linearity in the longwave cloud radiative effect results.

3.6 Longwave cloud radiative effect

The anthropogenic aerosol emissions exert a positive longwave cloud radiative effect ($\Delta CRE_{LW}$) of +0.65±0.02 W m$^{-2}$ (Fig. 11), partially offsetting the shortwave cloud radiative effect. Under the natural background assumption, the five largest contributing sources are East Asia, shipping, North America, Africa and the Middle East, and Southeast Asia. Under the present-day background assumption, only four sources make statistically significant contributions: shipping, North America, East Asia, and Southeast Asia.

Sub-linearity is evident in the $\Delta CRE_{LW}$ results. For every source region, the longwave warming effect is weaker under the present-day background assumption compared with under the natural background assumption. For example, East Asian emissions produce a longwave warming effect of +0.20±0.02 W m$^{-2}$ under the natural background assumption compared with a weaker effect of +0.06±0.02 W m$^{-2}$ under the present-day background assumption. $\Sigma_\Theta(All1-\Theta0)$ is only 23±11 % of $\Sigma_\Theta(\Theta1-All0)$. The $\Sigma_\Theta(All1-\Theta0)-\Sigma_\Theta(\Theta1-All0)$ difference in $\Delta CRE_{LW}$ (−0.91±0.17 W m$^{-2}$; Fig. 11) largely offsets the difference in $\Delta CRE_{SW}$ (+1.13±0.27 W m$^{-2}$; Fig. 8).

In contrast to the $\Delta CRE_{SW}$ global-mean results (Fig. 8), the relative contribution of each source region to the $\Delta CRE_{LW}$ summation depends on the assumption about background conditions (Fig. 11). For example, shipping emissions contribute 15±2 % under the natural background assumption, which is different from 29±16 % under the present-day background assumption. However, it should be noted that the change in the relative contribution is not statistically significant: the standard errors of the relative contributions are large, especially under the present-day background assumption.

With the exception of *All1–All0*, the zonal-mean $\Delta CRE_{LW}$ results are particularly noisy (Fig. S11). Having said that, the emissions from East Asia produce an interesting feature: under the natural background assumption, a positive peak occurs at approximately 20°N, near the southern end of the emissions region, while under the present-day background assumption, a negative peak occurs to the south, at approximately 10°N (Fig. S11a); a similar feature can be seen in the corresponding ice water results (Fig. S10a). However, the reason for this feature is unclear, and it may be due to noise.

**4 Sources of uncertainty**

Two major sources of uncertainty contribute to the results discussed above: uncertainty and variability in aerosol emissions; and model uncertainty. We briefly consider how each of these sources of uncertainty may influence the results.

4.1 Uncertainty and variability in aerosol emissions

Natural aerosol emissions are uncertain, contributing a large source of uncertainty to the effective radiative forcing of present-day anthropogenic aerosol emissions (Carslaw et al., 2013). In particular, emissions of dimethylsulfide (DMS) from phytoplankton are highly uncertain (Kettle & Andreae, 2000). The global total DMS emissions of 18.2 Tg(S) yr$^{-1}$ (Dentener et al., 2006) used in the present study are substantially lower than an alternative estimate of 28.1 Tg(S) yr$^{-1}$ (Lana et al., 2011). Due to the sub-linear response of clouds to aerosol emissions, an increase in DMS emissions would lead to a weakening of the cloud radiative effects of shipping emissions (Jin et al., 2018) and potentially other anthropogenic aerosol emissions.

Furthermore, despite our assumption that the natural and present-day aerosol emissions are stationary, aerosol emissions vary from year to year. For example, wildfire emissions – which we have categorized as a natural emissions source in the present study – are highly variable. When the wildfire emissions are averaged across different years, as is the case in the present study, this can lead to an overestimation of the strength of the cloud radiative effects of wildfires, again due to the sub-linear response of clouds to aerosol emissions (Grandey et al., 2016b). We expect that the interannual variability in wildfire emissions would also influence the cloud radiative effects of anthropogenic aerosol emissions.

Additionally, the "present-day" anthropogenic emissions themselves are not stationary. Aerosol emissions are continually changing due to changes in policy, technology, and demand for energy.

4.2 Model uncertainty

Even when emissions are held constant, there is widespread disagreement among global climate models as to the magnitude of the effective radiative forcing of anthropogenic aerosol emissions (Quaas et al., 2009; Shindell et al., 2013). Much of this model uncertainty can be attributed to model parameterizations relating directly to aerosol–cloud interactions, such as the aerosol activation scheme (Rothenberg et al., 2018). Assumptions about aerosol mixing state also contribute to the model uncertainty (Grandey et al., 2018a).

Available observational data have been unable to constrain tightly the model uncertainty associated with aerosol–cloud interactions. For ice clouds, experimental observations are unable to constrain the sign of the cloud radiative effect (Garimella et al., 2018). For liquid water clouds, differing approaches to improving observational constraints of aerosol–cloud interactions have been proposed (Mülmenstädt & Feingold, 2018). There is some evidence that many global climate models, including CESM-CAM5, may overestimate the response of cloud liquid water path to aerosol perturbations (Malavelle et al., 2017; Sato et al., 2018).

Model uncertainty contributes to uncertainty in the strength of the sub-linearity explored in this paper. For example, when investigating the radiative effects of co-varying shipping emissions and DMS emissions, Jin et al. (2018) found evidence of an even stronger sub-linear response of clouds to marine aerosol emissions when MAM3 is replaced by an alternative aerosol model (MARC, the two-Moment, Multi-Modal, Mixing-state-resolving Aerosol model for Research of Climate). We hypothesize that if we had used MARC in the present study, the difference in the cloud radiative effects between natural and present-day background conditions would be even stronger than in the MAM3 results presented here.

On the other hand, the anthropogenic aerosol effective radiative forcing produced by CAM5 is particularly strong compared with many other global climate models (Shindell et al., 2013). We hypothesize that if a model with weaker indirect effects were to be used, the absolute difference in the cloud radiative effects between natural and present-day background conditions would be weaker. We invite researchers with access to other global climate models to test this hypothesis.

**5 Discussion**

Aerosol–cloud–radiation interactions include non-linear processes. For example, the dependence of cloud droplet number concentration on aerosol availability and the dependence of cloud albedo on cloud optical thickness are both sub-linear (Ghan et al., 2013). In global climate models, relationships between CCN concentration and cloud properties are often non-linear (Gryspeerdt et al., 2017). Using CESM-CAM5, we have previously found evidence of sub-linearity in the shortwave cloud radiative effect of aerosol emissions from wildfires (Grandey et al., 2016b). In the current study, we find clear evidence of sub-linearity in the cloud radiative effects of anthropogenic aerosol emissions.

Due to the non-linear relationship between aerosol emissions and the cloud radiative effects, uncertainty in pre-industrial aerosol emissions contributes to uncertainty in the cloud radiative effects of present-day anthropogenic aerosol emissions. Using perturbed parameter simulations of a global aerosol model supplemented by emulators, Carslaw et al. (2013) found that natural emissions contribute a larger source of uncertainty to the net cloud radiative effect than anthropogenic emissions do. In general agreement with Carslaw et al. (2013), more recent studies have also emphasized that pre-industrial natural aerosol emissions contribute a large source of uncertainty to the cloud radiative effects (Carslaw et al., 2017; Regayre et al., 2018; Wilcox et al., 2015). Complementing these previous studies, we provide further evidence that background conditions influence estimated cloud radiative effects of anthropogenic aerosol emissions.

The global-mean cloud radiative effects of aerosol emissions depend on the source region of the emissions, as demonstrated by experiments using global climate models (Bellouin et al., 2016; Chen et al., 2018; Yang et al., 2017). For example, after normalizing by mass emitted, Bellouin et al. (2016) found that aerosol emissions from shipping and Europe likely exert a larger radiative effect than that exerted by emissions from East Asia, due to differing background conditions over the different regions. In the current study, we provide further evidence that the global-mean cloud radiative effects depend on the source region of the emissions: for example, shipping contributes 19–20% to anthropogenic $\Delta CRE_{SW}$ (Fig. 8), much larger than the contribution of only 11% to the anthropogenic sulfur dioxide emissions (Fig. 2).

The primary contribution of our study is found in the interplay between non-linearity, background conditions, and regional emissions. Just as pre-industrial aerosol emissions influence the estimated cloud radiative effects of anthropogenic aerosol emissions, emissions from other regions influence the cloud radiative effects of emissions from a target source region.

When investigating the radiative effects of aerosol emissions from a target source region, it is common practice to reduce emissions from the target source region while other emissions remain at their present-day values (Bellouin et al., 2016; Chen et al., 2018; Grandey et al., 2018b). In other words, present-day background conditions are often assumed. Although there is nothing inherently wrong with this assumption, our findings show that it is important to remember that such an assumption has been made and that this assumption influences the results. If natural background conditions were to be assumed, the estimated cloud radiative effects would likely be larger. In addition to the investigation of target source regions, this reasoning likely also applies to the investigation of specific aerosol species and target source sectors.

Finally, it should be noted that "present-day" background conditions are not stationary. The influence of aerosol emissions from different source regions will continue to evolve. This idea is explored briefly in the Conclusions below.

## 6 Conclusions

Due to transboundary transport, anthropogenic aerosol emissions from a given source region can influence aerosol abundance in other regions. Therefore, due to non-linearity in the response of clouds to aerosols, the radiative effects of aerosols emitted from one region may be influenced by emissions from other regions.

We have investigated the cloud radiative effects of anthropogenic aerosol emissions from different source regions under two different assumptions about background conditions: the "natural background" assumption, under which other anthropogenic emissions sources are absent; and the present-day background assumption, under which other anthropogenic emissions sources are present.

We have found that the shortwave and longwave cloud radiative effects are substantially weaker under present-day background conditions compared with under natural background conditions. For example, the shortwave cloud radiative effect of shipping is weaker under the present-day background assumption ($-0.39 \pm 0.03$ W m$^{-2}$) compared with under the natural background assumption ($-0.60 \pm 0.03$ W m$^{-2}$).

In light of these results, we recommend careful consideration of background conditions when quantifying the climate impacts of aerosol emissions from a given region.

We also note that "present-day" conditions are not stationary. The climate impacts of future aerosol emissions will depend on the geographic evolution of the emissions. For example, if aerosol emissions were to decrease across much of the world, this would move us closer to natural background conditions. Under such a scenario, the global-mean cloud radiative effects would weaken. However, if one region – or several regions – continue to emit at year-2000 emissions levels, the efficacy of these remaining emissions would increase: both the relative contribution and the absolute contribution of the source region to the cloud radiative effects would increase.

In this study, we have investigated the radiative effects of regional aerosol emissions using a model configuration with prescribed sea-surface temperatures. If corresponding coupled atmosphere–ocean simulations were to be performed, differing regional distributions of radiative effects would likely influence the climate response, including rainfall patterns (Chiang & Friedman, 2012; Grandey et al., 2016a; Wang, 2015). This provides an avenue for further research.

Both local and remote aerosol emissions influence air quality. Additionally, interaction between local and remote aerosol emissions influences climate.

**Code and Data Availability**

CESM 1.2.2 is available via http://www.cesm.ucar.edu/models/cesm1.2/. The scripts used to generate input data, configure the simulations, and produce the figures are available via https://github.com/grandey/draft2017a-region-rfp/, archived at https://doi.org/10.5281/zenodo.1987185. The simulation input data and the simulation output data analysed in this manuscript are archived at https://doi.org/10.6084/m9.figshare.6972827.

**Author Contributions**

BSG designed the experiment, configured the simulations, analyzed the data, produced the figures, and wrote the manuscript. CW provided supervisory guidance throughout the project, including contributions to the manuscript.

**Acknowledgments**

This research was supported by the National Research Foundation of Singapore under its Campus for Research Excellence and Technological Enterprise programme. The Center for Environmental Sensing and Modeling is an interdisciplinary research group of the Singapore-MIT Alliance for Research and Technology. This research was also supported by the U.S.

National Science Foundation (AGS-1339264) and the U.S. Department of Energy, Office of Science (DE-FG02-94ER61937). The CESM project is supported by the National Science Foundation and the Office of Science (BER) of the U.S. Department of Energy. The computational work for this research was performed on resources of the National Supercomputing Centre, Singapore (https://www.nscc.sg).

**Appendix A: Standard errors and statistical significance**

A.1 Calculating standard errors

For the mean of an individual scenario, e.g. *NAm1*, the standard error of the mean is calculated as

$$s_{\bar{x}} = \frac{s}{\sqrt{n}} \quad (1)$$

where *s* is the corrected sample standard deviation (calculated from the annual-means for different simulation years) and *n* is the number of simulation years analysed (60 for *All1* and *All0*; 30 for the *Θ1* and *Θ0* scenarios).

Standard errors are combined following Hogan (2006). For the difference between a pair of scenarios, e.g. *NAm1–All0*, the standard error of the difference between the scenarios is calculated by combining the standard errors of the means for the two scenarios:

$$s_{\bar{x}} = \sqrt{s_{\bar{x}1}^2 + s_{\bar{x}2}^2} \quad (2)$$

where $s_{\bar{x}1}$ is the standard error for the first scenario (*NAm1*) and $s_{\bar{x}2}$ is the standard error for the second scenario (*All0*), both calculated using Eq. (1).

For the fraction between a pair of scenarios, e.g. *NAm1/All0*, the standard error is calculated as

$$s_{\bar{x}} = \left|\frac{NAm1}{All0}\right| \times \sqrt{\left(\frac{s_{\bar{x}1}}{NAm1}\right)^2 + \left(\frac{s_{\bar{x}2}}{All0}\right)^2} \quad (3)$$

where $s_{\bar{x}1}$ is the standard error of *NAm1* and $s_{\bar{x}2}$ is the standard error of *All0*. Eq. (3) can also be used to calculate the percentage change between the pair of scenarios, given that (*NAm1 – All0*) / *All0* = (*NAm1 / All0*) – 1.

For a summation across multiple differences, e.g. $\Sigma_\Theta(\Theta 1–All0)$, it should be noted that the differences are not independent due to the repetition *All0*: $\Sigma_\Theta(\Theta 1–All0) = \Sigma_\Theta(\Theta 1) – 10 \times All0$. The standard error is calculated as

$$s_{\bar{x}} = \sqrt{(10 \times s_{\bar{x}1})^2 + \Sigma_\Theta(s_{\bar{x}\Theta}^2)} \quad (4)$$

where $s_{\bar{x}1}$ is the standard error for scenario *All0* and $s_{\bar{x}\Theta}$ is the standard error for scenario *Θ1*. The same approach applies for $\Sigma_\Theta(All1–\Theta 0)$.

For $\Sigma_\Theta(All1–\Theta 0)–\Sigma_\Theta(\Theta 1–All0)$ the standard error of the difference is calculated using Eq. (2) to combine the standard error of each summation.

For $\Sigma_\Theta(All1–\Theta 0)/\Sigma_\Theta(\Theta 1–All0)$, the standard error of the fraction is calculated as

$$s_{\bar{x}} = \left|\frac{\Sigma_\Theta(All1-\Theta 0)}{\Sigma_\Theta(\Theta 1-All0)}\right| \times \sqrt{\left(\frac{s_{\bar{x}1}}{\Sigma_\Theta(All1-\Theta 0)}\right)^2 + \left(\frac{s_{\bar{x}2}}{\Sigma_\Theta(\Theta 1-All0)}\right)^2} \quad (5)$$

where $s_{\bar{x}1}$ is the standard error of $\Sigma_\Theta(All1–\Theta 0)$ and $s_{\bar{x}2}$ is the standard error of $\Sigma_\Theta(\Theta 1–All0)$.

For the percentage contribution of a given difference, e.g. *NAm1–All0*, to a summation across multiple differences, e.g. $\Sigma_\Theta(\Theta1–All0)$, the standard error of the fraction is calculated similarly to Eq. (5):

$$s_{\bar{x}} = \left|\frac{NAm1-All0}{\Sigma_\Theta(\Theta1-All0)}\right| \times \sqrt{\left(\frac{s_{\bar{x}1}}{NAm1-All0}\right)^2 + \left(\frac{s_{\bar{x}2}}{\Sigma_\Theta(\Theta1-All0)}\right)^2} \times 100\% \qquad (6)$$

where $s_{\bar{x}1}$ is the standard error of *NAm1–All0* and $s_{\bar{x}2}$ is the standard error of $\Sigma_\Theta(\Theta1–All0)$. In applying Eq. (6), we assume that the numerator (*NAm1–All0*) and the denominator ($\Sigma_\Theta(\Theta1–All0)$) are independent: this simplification likely causes the standard error to be underestimated slightly.

For the percentage contributions shown in Figs. 4, S5, and S6, e.g. ((*NAm1–All0*)+(*All1–NAm0*))/(2×*All1*)×100%, the standard error of the fraction is calculated as

$$s_{\bar{x}} = \frac{1}{2} \times \left|\frac{(NAm1-All0)+(All1-NAm0)}{All1}\right| \times \sqrt{\left(\frac{s_{\bar{x}1}}{(NAm1-All0)+(All1-NAm0)}\right)^2 + \left(\frac{s_{\bar{x}2}}{All1}\right)^2} \times 100\% \qquad (7)$$

where $s_{\bar{x}1}$ is the standard error of (*NAm1–All0*)+(*All1–NAm0*) and $s_{\bar{x}2}$ is the standard error of *All1*. Again, we assume that the numerator and the denominator are independent, a simplification that likely causes the standard error to be underestimated slightly.

A.2 Testing statistical significance

For the difference between a pair of scenarios (e.g. *NAm1–All0*), significance is tested using an independent two-sample *t*-test, with the assumption that both distributions have equal variance. Annual-mean data for each simulation year are used as the input. Each sample contains 30 or 60 simulation years, so we assume that the central limit theorem can be applied. For the bar charts showing global-mean results (e.g. Fig. 3), a two-tailed p-value threshold of 0.01 is used. For the maps shown in Figs. S19–S21, the false discovery rate is controlled (Benjamini & Hochberg, 1995; Wilks, 2016), using a significance level of 0.05.

For other combinations of scenarios, e.g. $\Sigma_\Theta(\Theta1–All0)$, significance is tested using the 99% confidence interval. Assuming that the data are normally distributed, the 99% confidence interval is calculated by multiplying the standard error by a factor of 2.576, providing the bounds above and below the mean. This method is used for the twelfth and final bars of the bar charts.

For the maps showing percentage contributions (e.g. Fig. 4), significance is tested using the 99% confidence interval for each grid box. Due to the complication of using confidence intervals, we do not control the false discovery rate. Therefore, additional caution is required when interpreting these results: in the absence of a genuine signal, the expected rate of occurrence of false positives is 1%. To provide additional guidance, the percentage occurrence of statistically significant negative and positive values is provided underneath each map; area-weighting is not applied when calculating the percentage occurrence of significant results.

**Appendix B: Interpretation of bar charts**

Many of the figures are bar charts showing global-mean results (e.g. Fig. 3). The unconventional arrangement of the bars has been carefully chosen to aid interpretation of the results. Because the discussion of the results depends heavily on these bar charts, we recommend that readers carefully study one of these bar charts (e.g. Fig. 3). When readers are familiar with the interpretation of one of these bar charts, it should be relatively easy to interpret the other bar charts.

The first bar shows the difference between the *All1* simulation and the *All0* simulation for a given variable of interest (e.g. sulfate aerosol burden). This *All1–All0* difference reveals the total impact of year-2000 anthropogenic aerosol emissions.

The second to eleventh bars show differences between members of the $\Theta 1$ group of simulations and the *All0* simulation. For example, the second bar of Fig. 3 shows the *NAm1–All0* difference, revealing the impact of North American anthropogenic aerosol emissions under the natural background assumption. The bars are both staggered and stacked, such that the bottom of each consecutive bar is located at the top of the previous bar. The bars are ordered in either descending (e.g. Fig. 3) or ascending (e.g. Fig. 8) order.

The twelfth bar shows the sum of the second to eleventh bars, $\Sigma_\Theta(\Theta 1$–*All0)*, indicating the sum of the impacts modelled by the $\Theta 1$ group of simulations with respect to All0. Green text indicates the percentage contribution of a given bar to the thirteenth bar. For example, the second bar of Fig. 3, showing *NAm1–All0*, is 18±0 % of the height of the twelfth bar: this indicates that the North American emissions contribute 18% to the combined anthropogenic sulfate burden modelled by the $\Theta 1$ group of simulations, the largest contribution from any source region.

The thirteenth to twenty-second bars show differences between the All1 simulation and members of the $\Theta 0$ group of simulations. For example, the thirteenth bar of Fig. 3 shows the *All1–NAm0* difference, revealing the impact of North American anthropogenic aerosol emissions under the present-day background assumption.

The final bar shows the sum of the thirteenth to twenty-second bars, $\Sigma_\Theta(All1$–$\Theta 0)$, indicating the sum of the impacts modelled by the $\Theta 0$ group of simulations with respect to *All1*. As before, green text indicates the percentage contributions to the final bar.

It is important to note that the first, twelfth, and final bars are calculated differently, allowing non-linearity to be investigated. *All1–All0* (first bar) represents the "true" total impact of anthropogenic aerosol emissions in the model; $\Sigma_\Theta(\Theta 1$–*All0)* (twelfth bar) and $\Sigma_\Theta(All1$–$\Theta 0)$ (final bar) provide alternative estimates of the total impact of anthropogenic aerosol emissions, revealing the total impact diagnosed when emissions are switched on and off for different source regions in turn, under different assumptions about background conditions.

For some of the bar charts (e.g. Fig. 3), *All1–All0*, $\Sigma_\Theta(\Theta 1$–*All0)*, and $\Sigma_\Theta(All1$–$\Theta 0)$ are very similar, as would be expected for a linear system. However, for other bar charts (e.g. Fig. 8), differences between *All1–All0*, $\Sigma_\Theta(\Theta 1$–*All0)*, and $\Sigma_\Theta(All1$–$\Theta 0)$ indicate non-linearity. For example, the difference between $\Sigma_\Theta(\Theta 1$–*All0)* and $\Sigma_\Theta(All1$–$\Theta 0)$ in Fig. 8 reveals the sub-linear influence of anthropogenic aerosol emissions on the shortwave cloud radiative effect. When the magnitude of $\Sigma_\Theta(All1$–$\Theta 0)$ is smaller than the magnitude of $\Sigma_\Theta(\Theta 1$–*All0)*, this implies sub-linearity. The same applies to the impact of emissions from a specific source region, $\Theta$: if the magnitude of *All1–$\Theta 0$* is smaller than the magnitude of *$\Theta 1$–All0*, this implies sub-linearity, namely that the emissions from region $\Theta$ exert a weaker influence if the background conditions include present-day emissions from other regions.

**Figures**

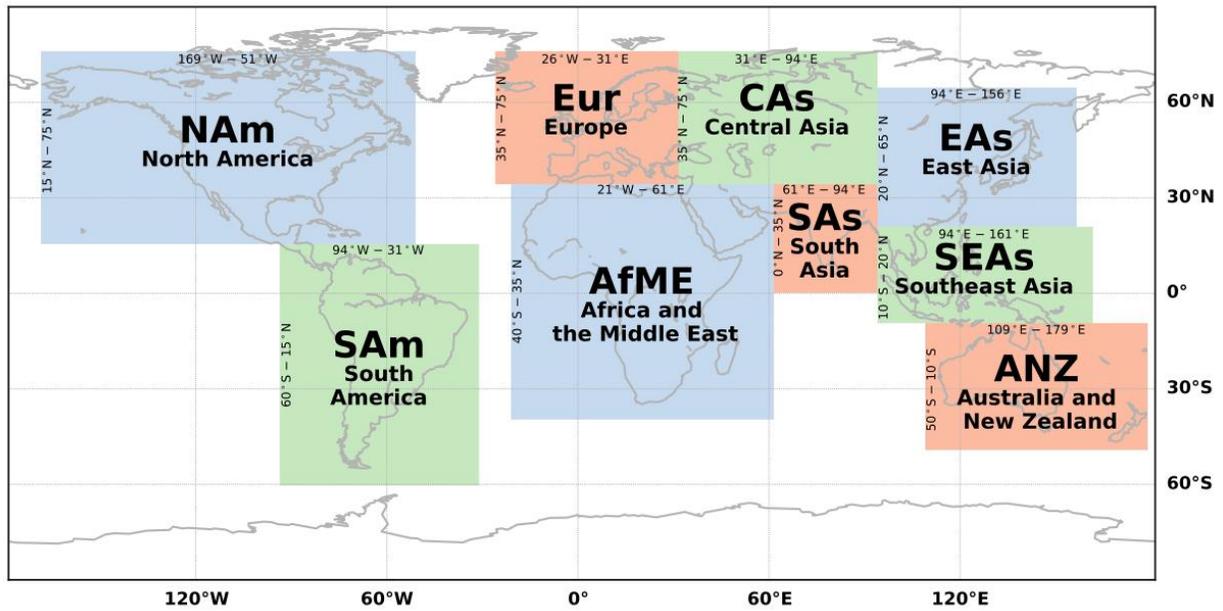

**Figure 1.** Source regions investigated in this study. Region bounds are written at the northern and western edges of each rectangle.

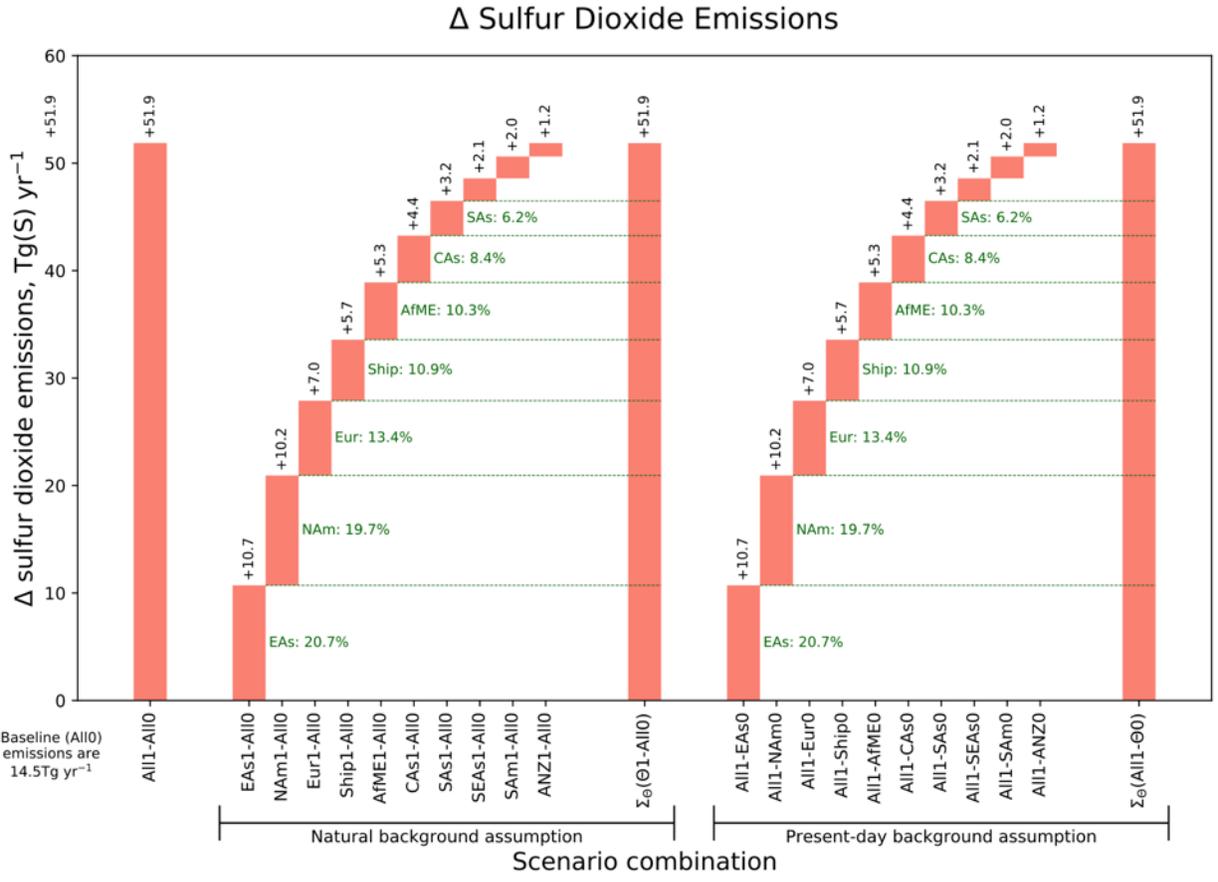

**Figure 2.** Differences in global emissions of sulfur dioxide (including primary sulfate) for different combinations of scenarios. Approximately 2.5% of the sulfur dioxide is emitted as primary sulfate aerosol (Text S1). The green text shows the percentage contribution of each source region to $\Sigma_\Theta(All1-\Theta 0)$ and $\Sigma_\Theta(\Theta 1-All0)$. The natural baseline emissions of *All0* are written at the bottom left of the figure.

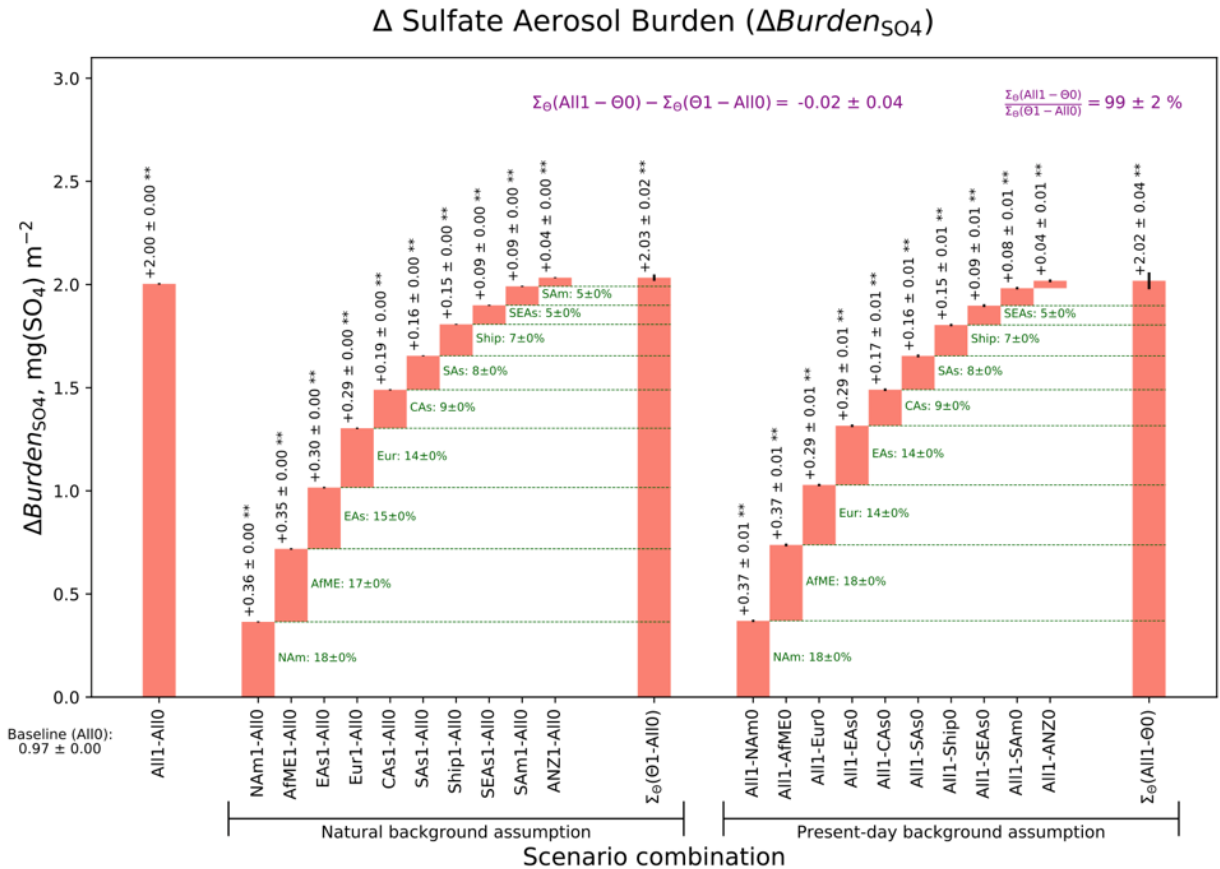

**Figure 3.** Differences in global-mean annual-mean sulfate aerosol burden for different combinations of scenarios. The global-means are calculated using both land and ocean data. Area-weighting is applied. The text above each bar shows the difference, standard error, and an indication of statistical significance: black text and '**' indicates statistical significance at a significance level of 0.01; grey text would indicate insignificance (as is the case for three bars in Fig. 6). The green text shows the percentage contribution of each source region to $\Sigma_\Theta(\Theta 1-All0)$ and $\Sigma_\Theta(All1-\Theta 0)$, with associated standard errors. The purple text shows $\Sigma_\Theta(All1-\Theta 0)-\Sigma_\Theta(\Theta 1-All0)$ and $\Sigma_\Theta(All1-\Theta 0)/\Sigma_\Theta(\Theta 1-All0)$, with associated standard errors. The lack of '**' following the $\Sigma_\Theta(All1-\Theta 0)-\Sigma_\Theta(\Theta 1-All0)$ difference indicates that the difference is statistically insignificant; the significance of $\Sigma_\Theta(All1-\Theta 0)/\Sigma_\Theta(\Theta 1-All0)$ is not determined. The natural baseline result of *All0* is written at the bottom left of the figure. See Appendix A for an explanation of how standard errors are calculated and how statistical significance is tested. See Appendix B and for an explanation of how to interpret the figure components.

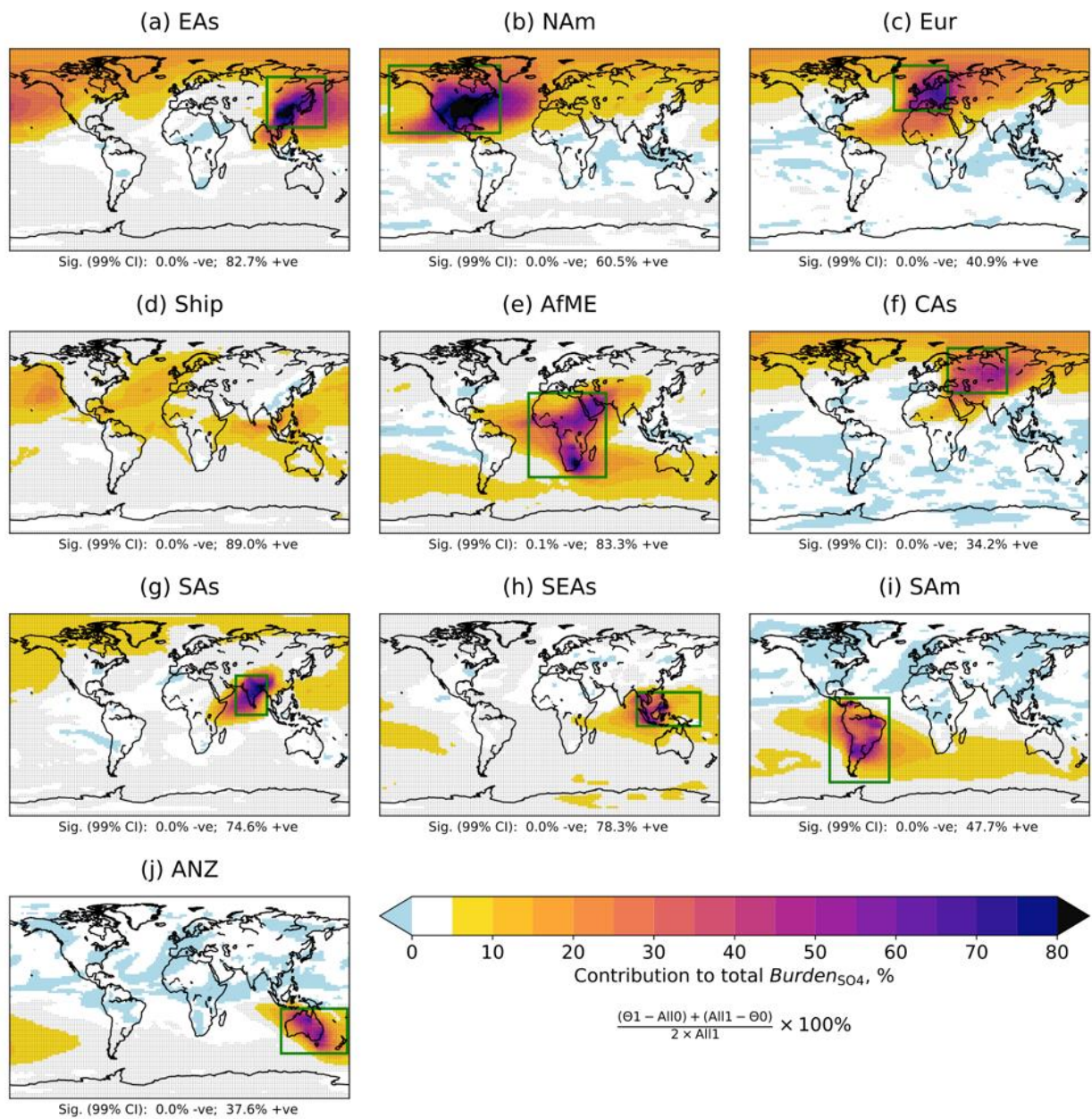

**Figure 4.** Percentage contributions of each source region to the total sulfate aerosol burden. The total burden includes contributions from natural emissions sources. The percentage contributions are calculated using the equation below the colour bar, where $\Theta$ is the source region. Stippling indicates statistical significance, calculated using the 99% confidence interval (Appendix A). The percentage occurrence of statistically significant negative and positive values is written underneath each map. With the exception of shipping (panel d), the longitude and latitude bounds of each source region are indicated by a green rectangle.

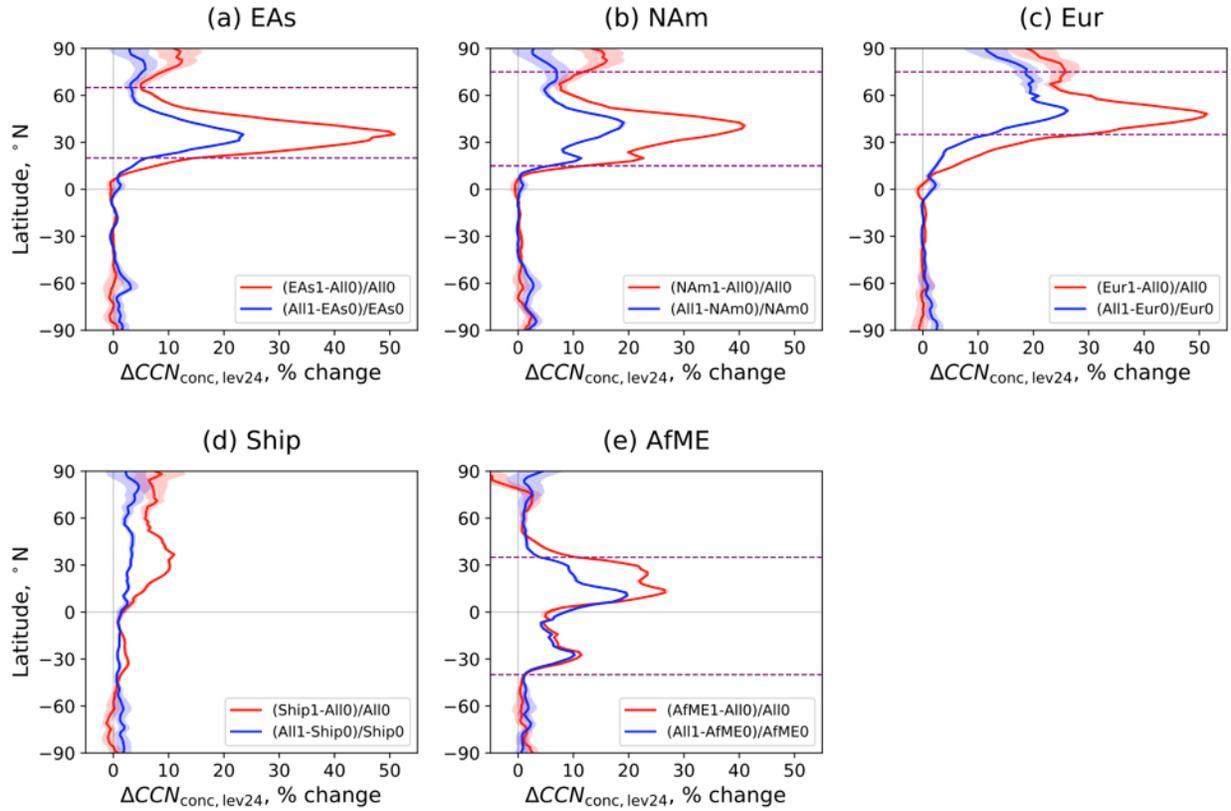

**Figure 5.** Percentage changes in zonal-mean annual-mean CCN concentration at a supersaturation of 0.1% in model level 24 (~860hPa, in the lower troposphere) for different combinations of scenarios. Shading indicates standard errors (Eq. (3) in Appendix A). The five source regions have been selected because they have the largest sulfur dioxide emissions. In panels (a)–(c) and (e), purple dashed lines indicate the latitude bounds of the respective source region.

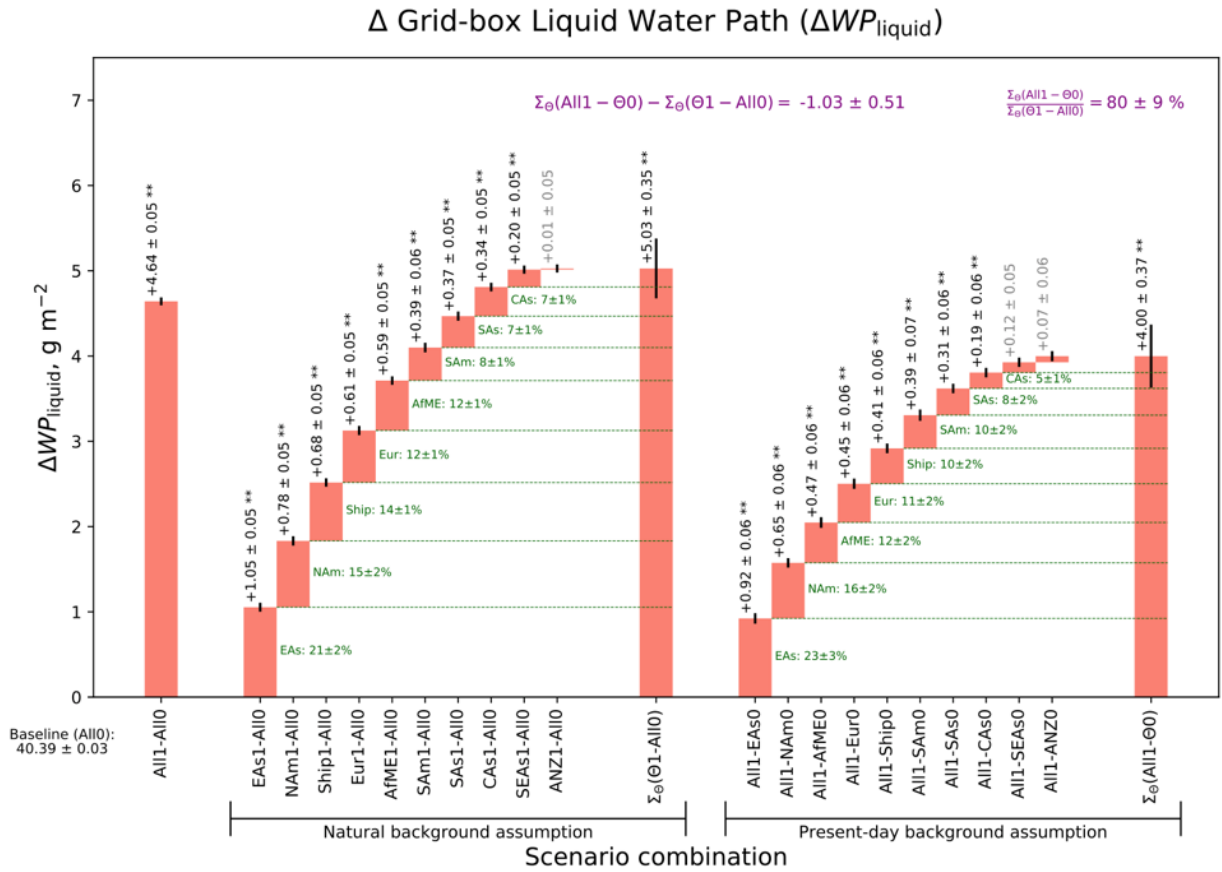

**Figure 6.** Differences in global-mean annual-mean grid-box liquid water path for different combinations of scenarios. See the Fig. 3 caption for an explanation of the figure components.

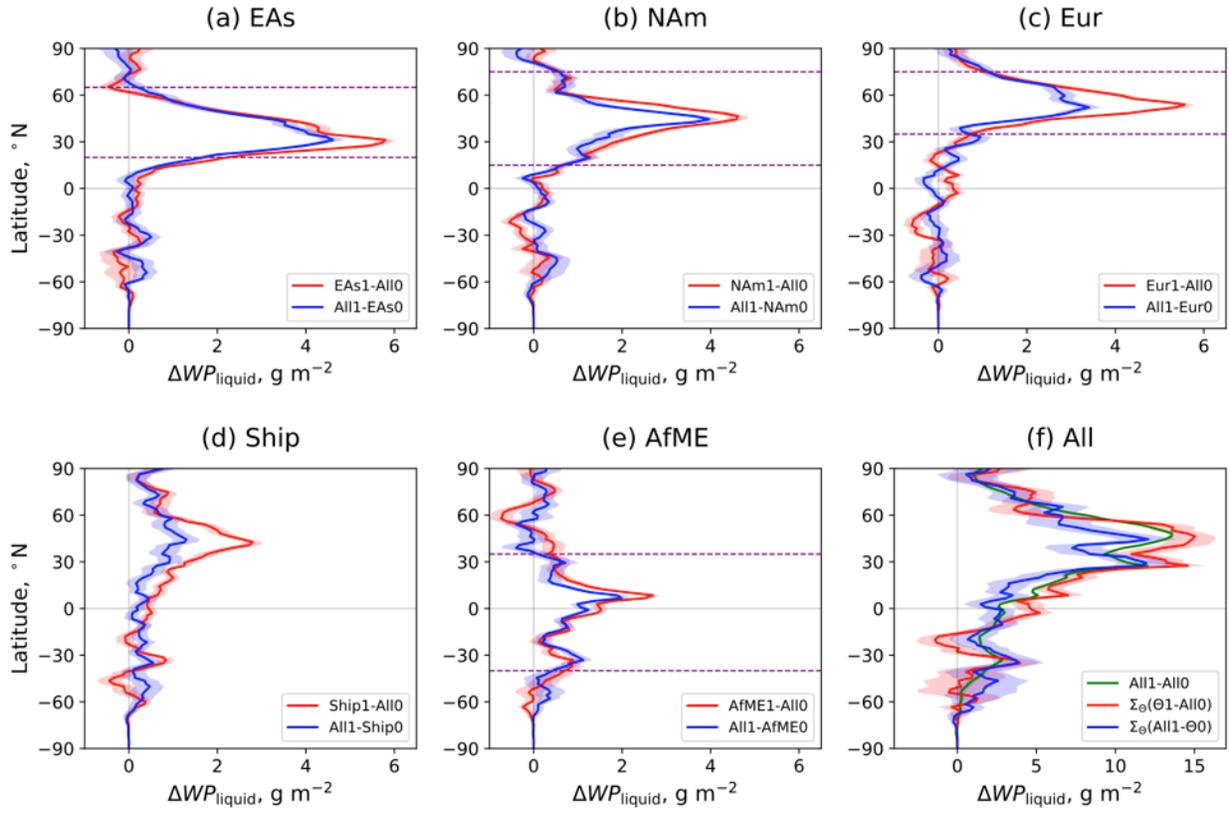

**Figure 7.** Differences in zonal-mean annual-mean grid-box liquid water path for different combinations of scenarios. Shading indicates standard errors (Appendix A). In panels (a)–(e), the five source regions have been selected because they have the largest sulfur dioxide emissions. In panels (a)–(c) and (e), purple dashed lines indicate the latitude bounds of the respective source region.

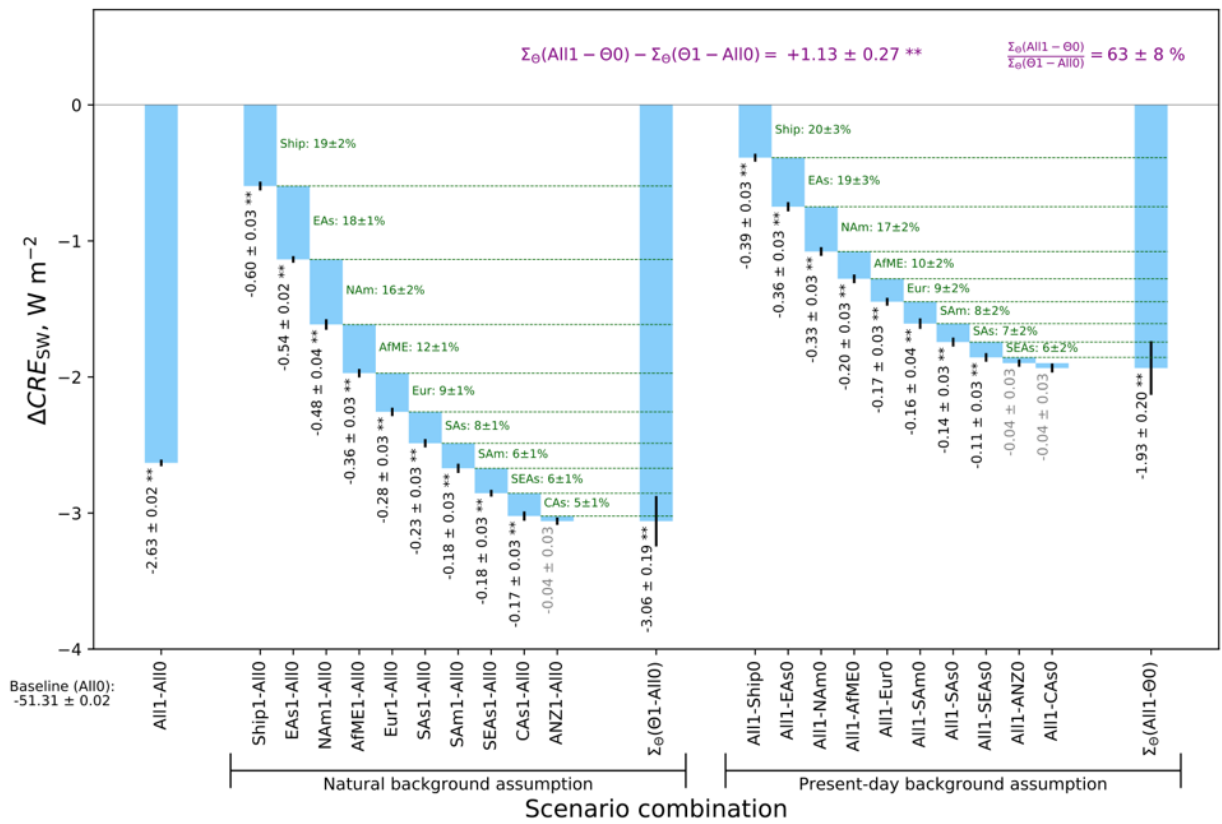

**Figure 8.** Differences in global-mean annual-mean clean-sky shortwave cloud radiative effect for different combinations of scenarios. See the Fig. 3 caption for an explanation of the figure components.

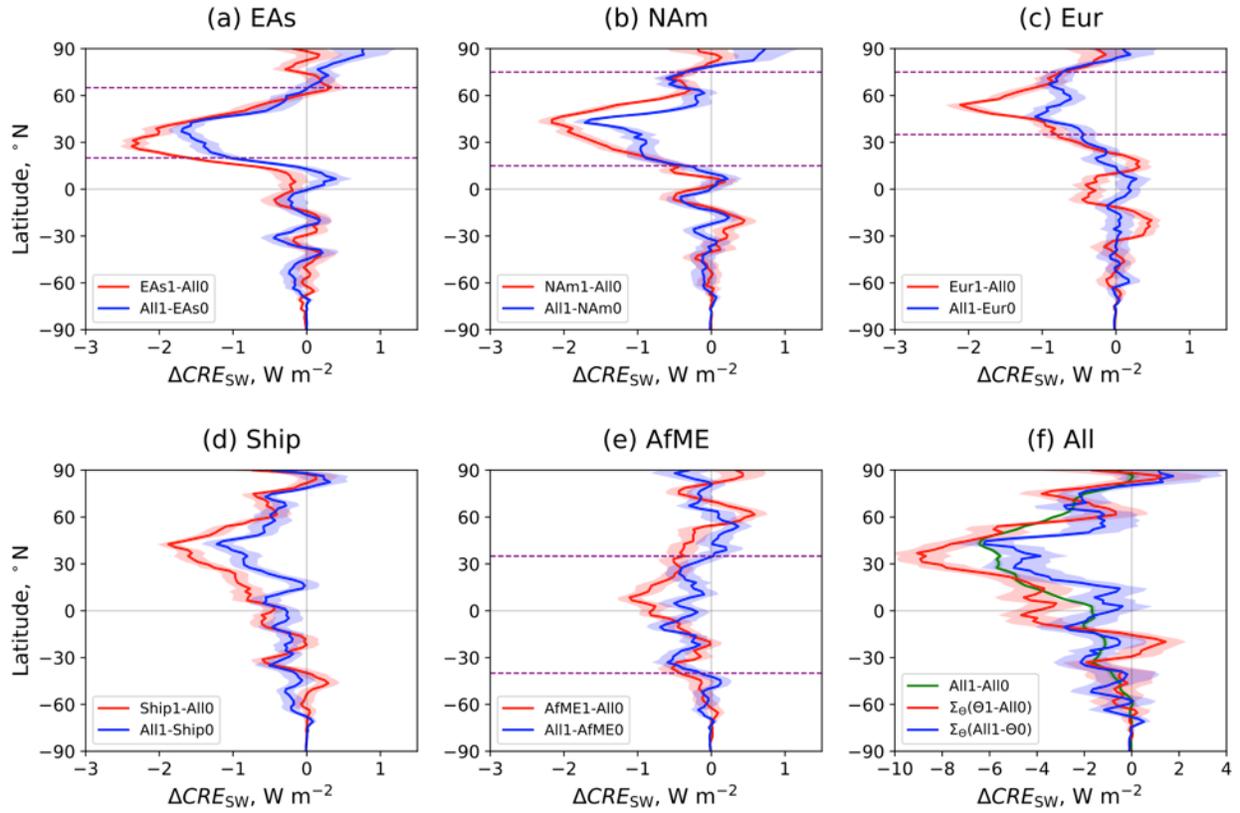

**Figure 9.** Differences in zonal-mean annual-mean clean-sky shortwave cloud radiative effect for different combinations of scenarios. See the Fig. 7 caption for an explanation of the figure components.

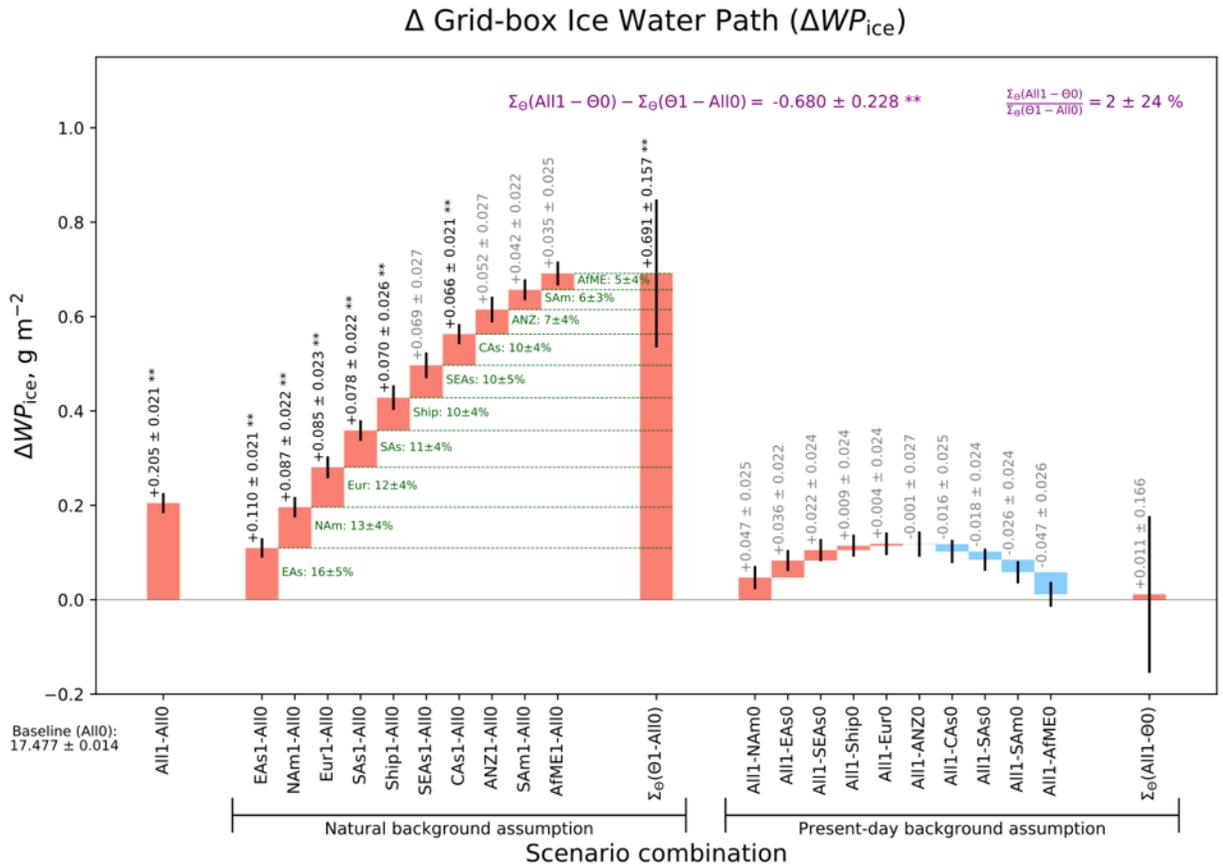

**Figure 10.** Differences in global-mean annual-mean grid-box ice water path for different combinations of scenarios. See the Fig. 3 caption for an explanation of the figure components.

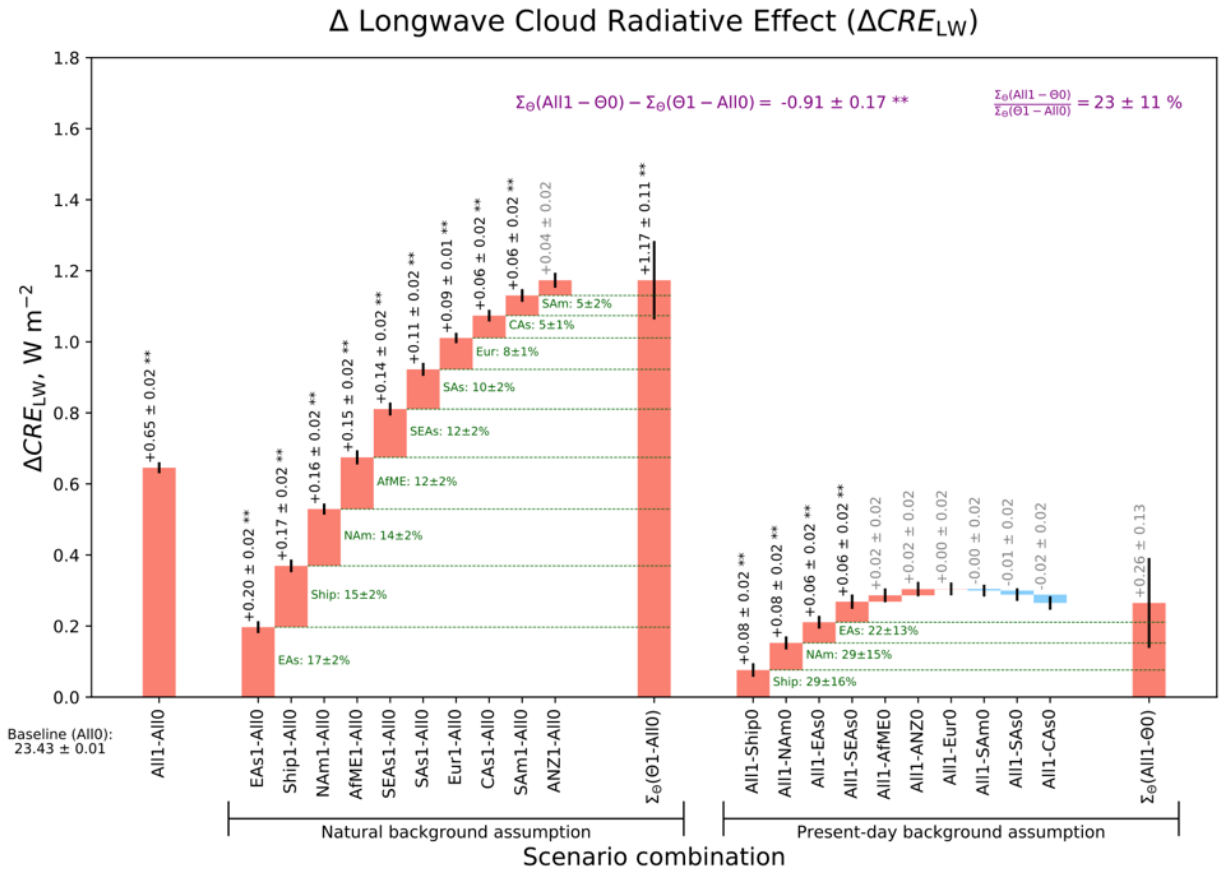

**Figure 11.** Differences in global-mean annual-mean longwave cloud radiative effect for different combinations of scenarios. See the Fig. 3 caption for an explanation of the figure components.

Supporting Information for

**Background Conditions Influence the Estimated Cloud Radiative Effects of Anthropogenic Aerosol Emissions from Different Source Regions**

Benjamin S. Grandey[1], and Chien Wang[2,1]

[1]Center for Environmental Sensing and Modeling, Singapore-MIT Alliance for Research and Technology, Singapore, Singapore.

[2]Center for Global Change Science, Massachusetts Institute of Technology, Cambridge, Massachusetts, USA.

**Contents of this file**

    Introduction
    Text S1: Primary sulfate aerosol emissions
    Figures S1 to S21
    References

**Introduction**

This file contains supplementary text and figures. The scenarios and model configuration are described in the main manuscript. The analysis code is available via https://github.com/grandey/draft2017a-region-rfp, archived at https://doi.org/10.5281/zenodo.1987185.

Readers interested in regional analysis of the net radiative effect are referred to presentation slides archived at https://doi.org/10.6084/m9.figshare.7380191.

**Text S1: Primary sulfate aerosol emissions**

As described below, the configuration files for all simulations contained a minor bug that affects the primary sulfate emissions. However, as demonstrated below, the bug does not substantially impact any of the results presented in this study.

**Description of minor bug**

It is assumed that 2.5% of the sulfur dioxide is emitted as primary sulfate (Dentener et al., 2006; Liu et al., 2012). The mode (either accumulation or Aitken) and vertical level (either surface or elevated) of the primary sulfate emissions depends on the source sector (Fig. S12). The emitted size distributions, described in the Supplement of Liu et al. (2012), are specified via number emissions files that accompany the mass emissions files.

However, the configuration files for the simulations described in the manuscript contained a bug: the line specifying the primary sulfate Aitkin-mode surface mass emissions file incorrectly pointed to the corresponding accumulation-mode surface mass emissions file (Fig. S12). This has three conquences:

1. The domestic and transport primary sulfate mass emissions, contained in the Aitkin-mode surface emissions files, are missing from the simulations.

2. The shipping, waste treatment, and agricultural waste burning primary sulfate mass emissions, contained in the accumulation-mode surface emissions files, are twice what they should be. The erroneous additional emissions are emitted in the Aitkin-mode.

3. The erroneous Aitkin-mode surface emissions have incorrect size distributions, because the Aitkin-mode surface number emissions still follow the number emissions derived for the domestic and transport sectors.

The simulations are still consistent with the intended scenarios. For example, *Ship0* still contains no shipping emissions.

The organic carbon aerosol and black carbon aerosol emissions are not affected by the bug. The bug only affects the primary sulfate aerosol emissions.

**Impact of bug on sulfur emissions**

The impact of the bug on the global primary sulfate and total sulfur emissions is summarised in Fig. S13. The bug does not affect scenario *All0*, because the bug affects only anthropogenic emissions sources.

For scenario *All1*, the bug leads to an error of $-0.015$ Tg(S) yr$^{-1}$ globally, only $-0.02\%$ of the total sulfur emissions (Fig. S13). Over land, the bug generally leads to a slight decrease in the sulfur emissions (Figs. S13 and S14b), due to the missing primary sulfate emissions from the domestic and transport sectors. Over ocean, the bug generally leads to a slight increase in the sulfur emissions, due to the erroneously doubled primary sulfate emissions from shipping. The error introduced by the bug is limited by the fact that only a small fraction of the sulfur dioxide is emitted as primary sulfate. Hence, in any given location, the magnitude of the error is at most 2.5% of the total sulfur emissions, and is generally much less than this (Fig. S14c).

For scenario *Ship1*, the bug generally leads to a slight increase in the sulfur emissions over the ocean (Figs. S13 and S15b), as was the case for *All1*. However, because scenario *Ship1* does not include any land-based anthropogenic emissions source, the bug does not affect the emissions over inland areas. When summed globally, the increase in the emissions over ocean leads to a total error of +0.138 Tg(S) yr$^{-1}$, +0.36% of the total sulfur emissions – this is the largest relative error in the the global emissions found for any scenario.

For the other members of the *Θ1* group of scenarios, such as *EAs1*, the bug generally leads to a slight decrease in the sulfur emissions over the target land region: for example, for scenario *EAs1*, the bug leads to a decrease in the sulfur emissions over East Asia (Fig S16b). The bug does not affect the sulfur emissions over other land regions and over the open ocean.

For scenario *Ship0*, the bug generally leads to a slight decrease in the sulfur emissions over land (Fig. S17b), as was the case for *All1*. However, because scenario *Ship0* does not include any shipping emissions, the bug does not affect the emissions over open ocean. When summed globally, the decease in the emissions over land leads to a total error of −0.154 Tg(S) yr$^{-1}$ – this is the largest absolute error in the global emissions found for any scenario.

For the other members of the *Θ0* group of scenarios, the bug does not affect emissions over the target land region: for example, for scenario *EAs0*, the bug does not affect the sulfur emissions over inland areas in East Asia (Fig. S18b). Over other land regions and over the open ocean, the error in the emissions is the same as for *All1*: a slight decrease over land, and a slight increase over ocean.

To summarise, for any given scenario in the *Θ1* and *Θ0* groups of scenarios, the absolute error in the emissions at a given location over the open ocean or inland is either (i) zero or (ii) the same as scenario *All1*. (Over coastal areas, the error might differ due to contributions from both shipping and anthropogenic land-based sources.) For example, *Ship1* has the same emissions error as *All1* over the open ocean, while *Ship0* has the same emissions error as *All1* over inland areas. At any given location, the magnitude of the error is at most 2.5% of the total sulfur emissions, and is generally much less than this.

**Impact of bug on results**

As explained in the paragraphs above, the bug has only a small impact on the total sulfur emissions for any given scenario. Hence, it is to be expected that the bug will not substantially impact the results. However, to confirm this assertion, we performed an additional simulation: scenario *Correct1*, which uses the bug-free emissions intended for scenario *All1*. (*Correct1* also has an analysis period of 60 years.) Comparison of *Correct1* with *All1*, and *Correct1*–*All0* with *All1*–*All0*, reveals the impact of the bug on the model results.

It could be argued that corrected versions of *Ship1* and *Ship0* should also be investigated, because *Ship1* and *Ship0* have the largest errors in the global sulfate emissions. However, at any given location, the absolute error in the sulfur emissions of *Ship1* or *Ship0* is no larger than that of *All1*. Because sulfate aerosol has a relatively short lifetime in the atmosphere, and hence is not transported far from the emissions source, it is to be

Supporting Information

expected that over ocean the impact of the bug on the *Ship1* results should be similar to the impact on the *All1* results. Correspondingly, over land the impact of the bug on the *Ship0* results should be similar the impact on the *All1* results. Hence we argue that investigation of *Correct1* is sufficient to establish an upper bound of the impact of the bug on all other scenarios.

We apply two approaches in order to compare the results of *Correct1* with *All1*. First, maps of the *All1–Correct1* differences in the sulfate aerosol burden and cloud radiative effects are shown in Figs. S19–S21. These maps reveal that the differences between *All1* and *Correct1* are generally very small and statistically insignificant over both land and ocean – this also constrains the impact of the bug on the other scenarios, as argued in the paragraph above.

Second, global-mean results are shown underneath each map in Figs. S19–S21. In all three cases, the *All1–Correct1* difference is statistically insignificant, with zero falling within the bounds of the standard error.

**Summary**

To summarise, (i) the bug in the primary sulfate emissions has only a small effect on the total sulfur emissions, and (ii) the bug does not significantly impact the model results. Therefore, we conclude that the bug in the primary sulfate emissions does not meaningfully impact any of the results or conclusions presented in the manuscript.

**Figures S1 to S21**

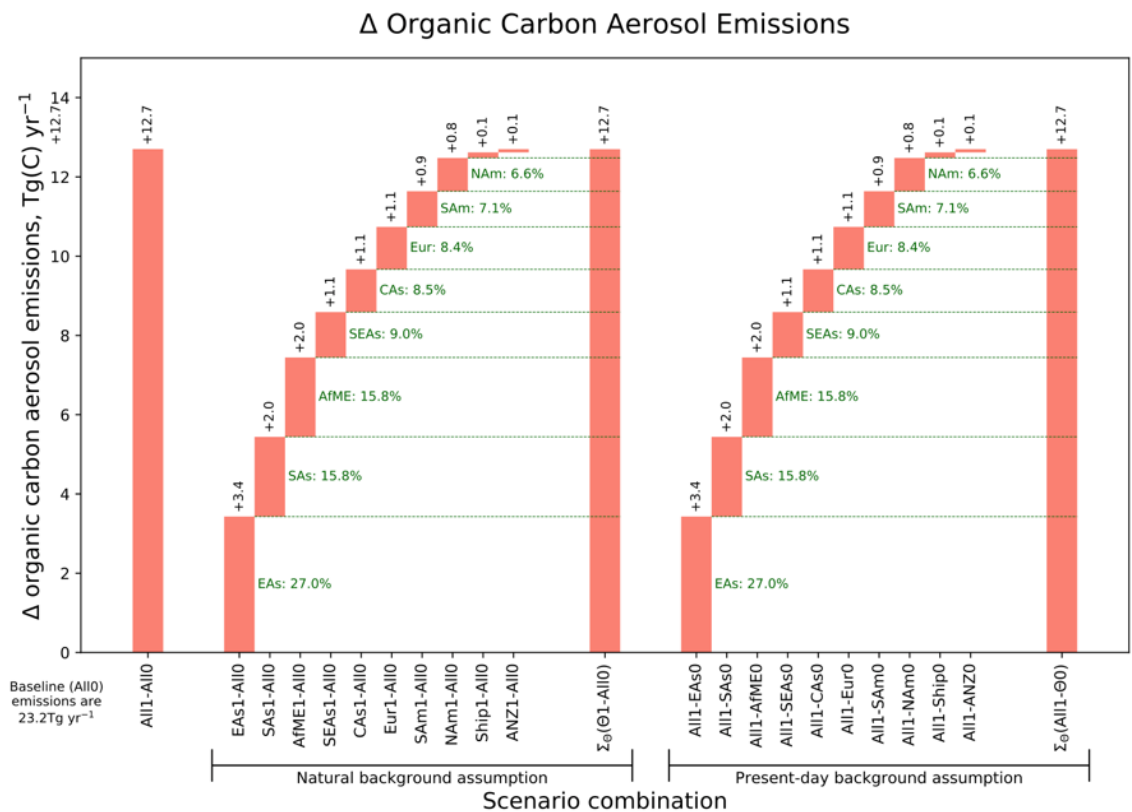

**Figure S1.** Differences in global emissions of primary organic carbon aerosol for different combinations of scenarios. See the Fig. 2 caption for an explanation of the figure components.

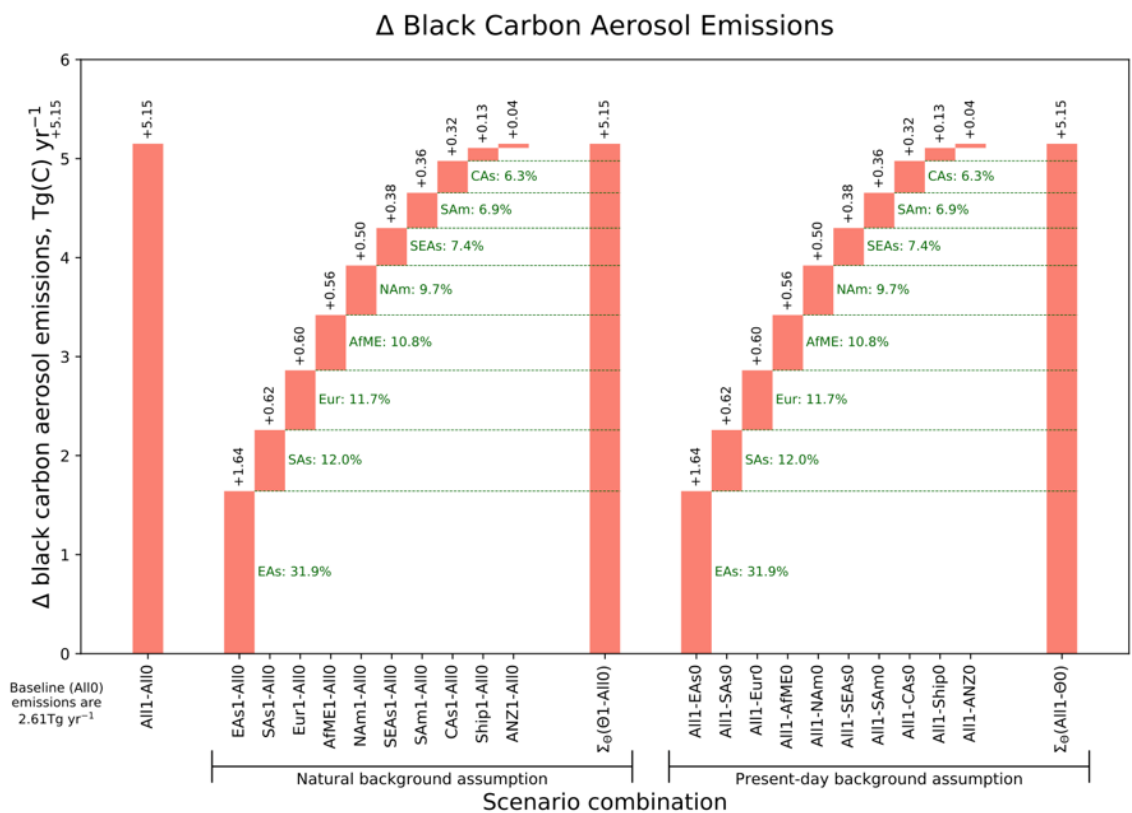

**Figure S2.** Differences in global emissions of black carbon aerosol for different combinations of scenarios. See the Fig. 2 caption for an explanation of the figure components.

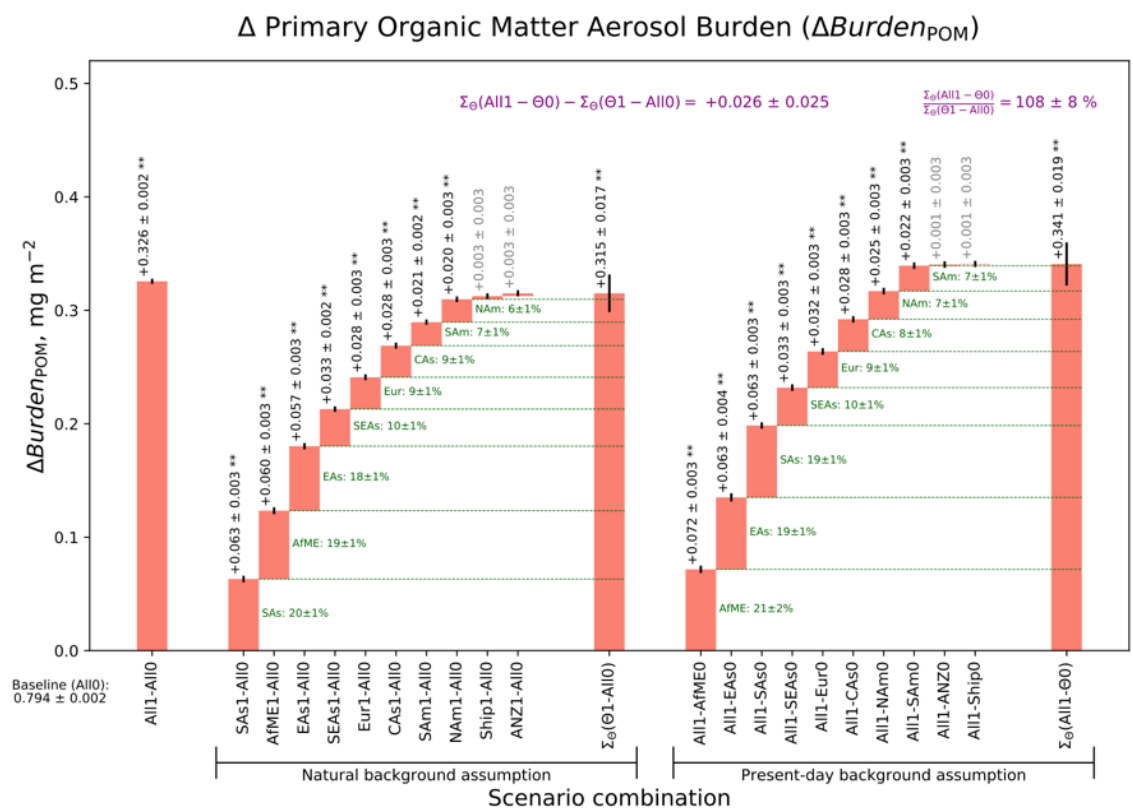

**Figure S3.** Differences in annual-mean global-mean primary organic matter aerosol burden for different combinations of scenarios. See the Fig. 3 caption for an explanation of the figure components.

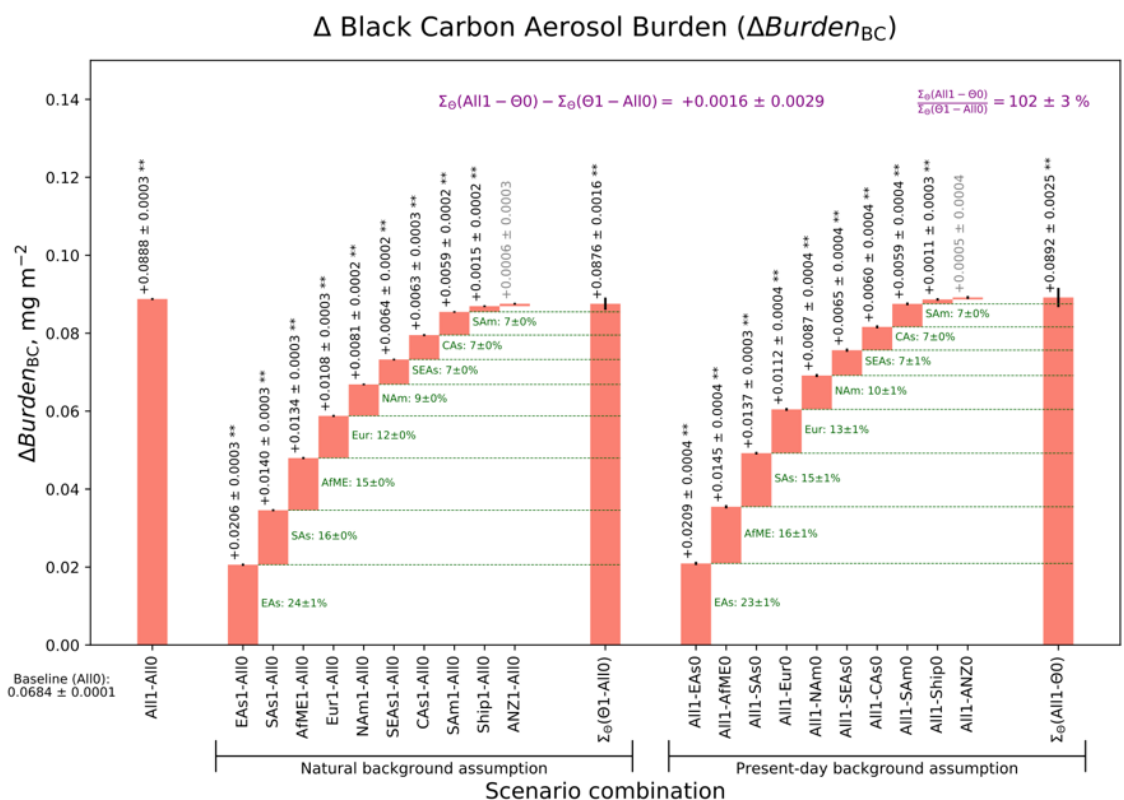

**Figure S4.** Differences in annual-mean global-mean black carbon aerosol burden for different combinations of scenarios. See the Fig. 3 caption for an explanation of the figure components.

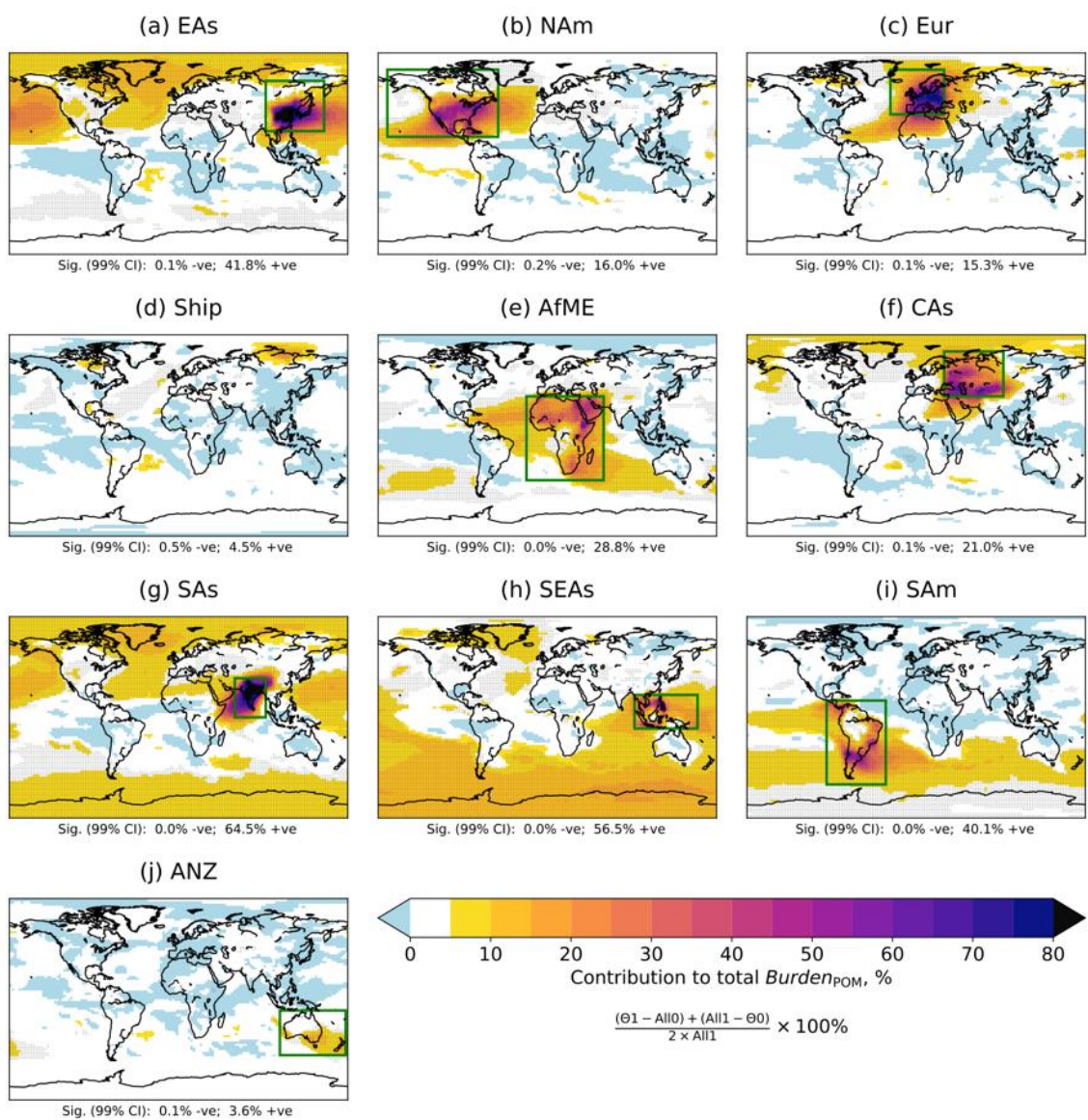

**Figure S5.** Percentage contributions of each source region to the annual-mean total primary organic matter aerosol burden. See the Fig. 4 caption for an explanation of the figure components.

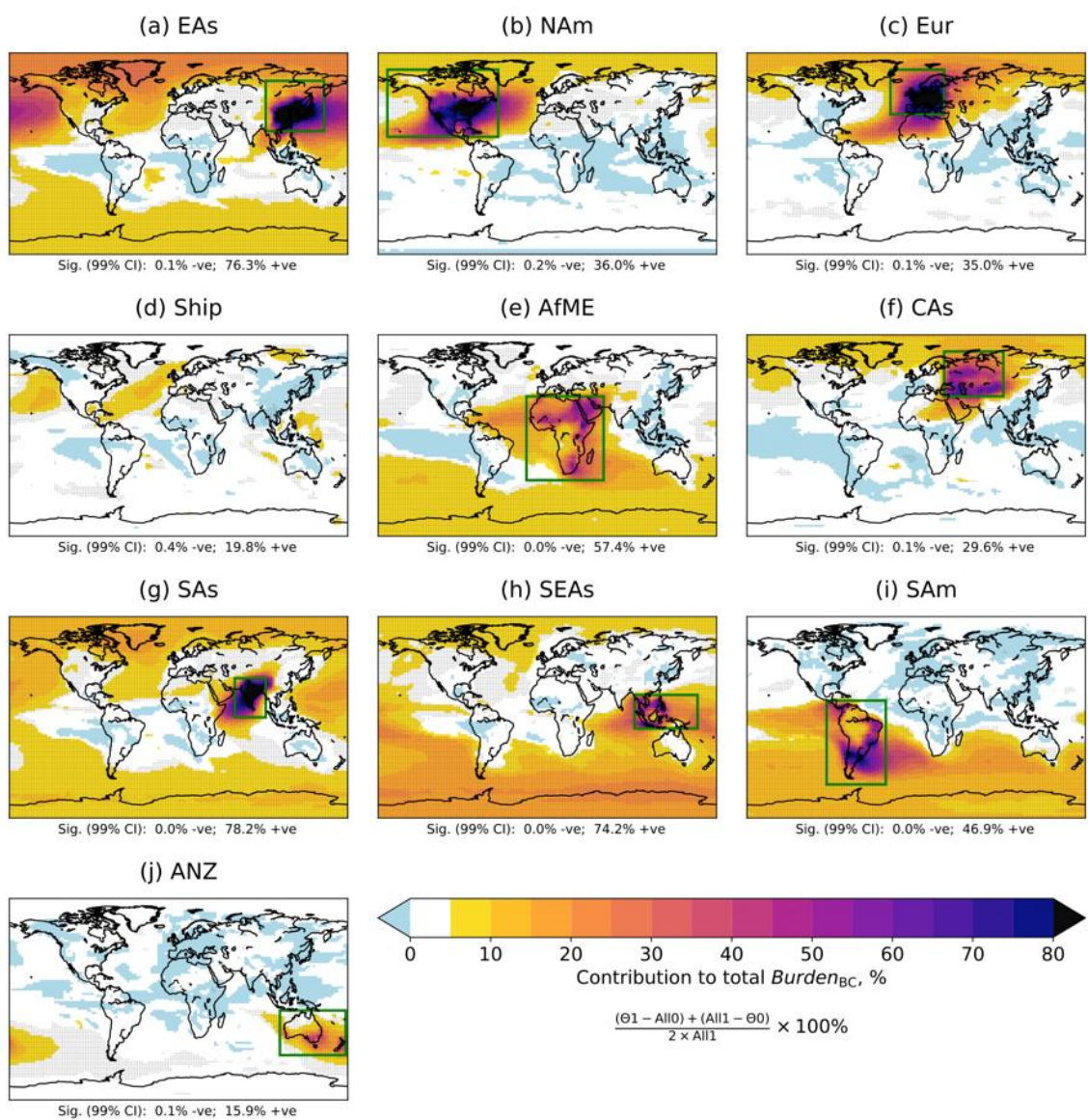

**Figure S6.** Percentage contributions of each source region to the annual-mean total black carbon aerosol burden. See the Fig. 4 caption for an explanation of the figure components.

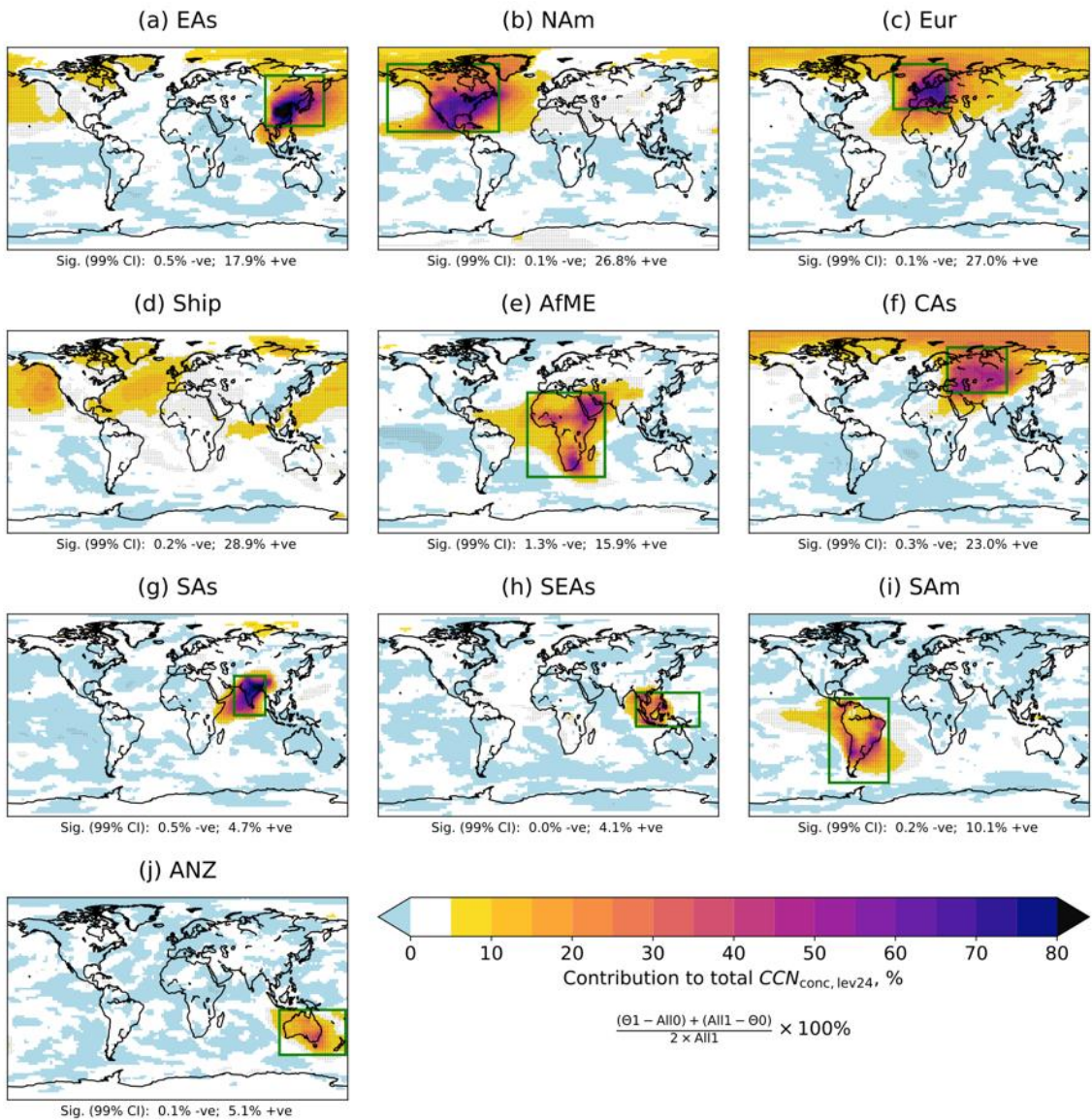

**Figure S7.** Percentage contributions of each source region to the annual-mean total CCN concentration at a supersaturation of 0.1% in model level 24 (~860hPa, in the lower troposphere). See the Fig. 4 caption for an explanation of the figure components.

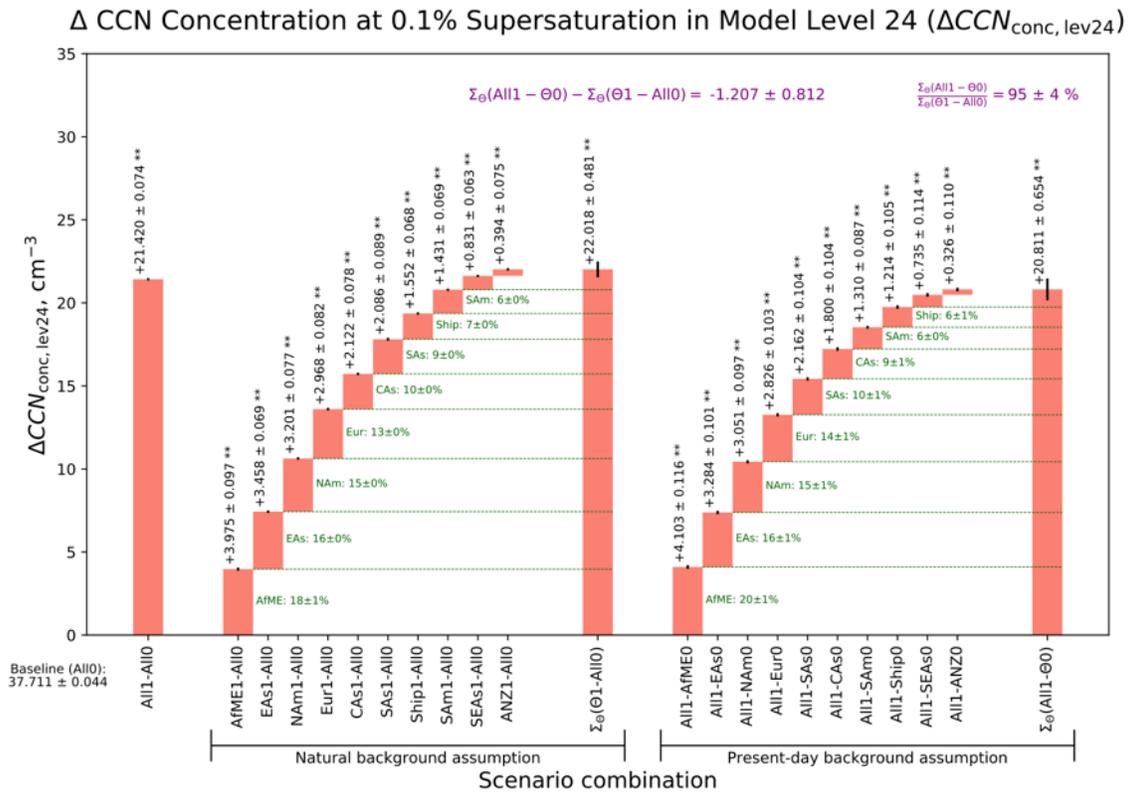

**Figure S8.** Differences in annual-mean global-mean CCN concentration at a supersaturation of 0.1% in model level 24 (~860hPa, in the lower troposphere) for different combinations of scenarios. See the Fig. 3 caption for an explanation of the figure components.

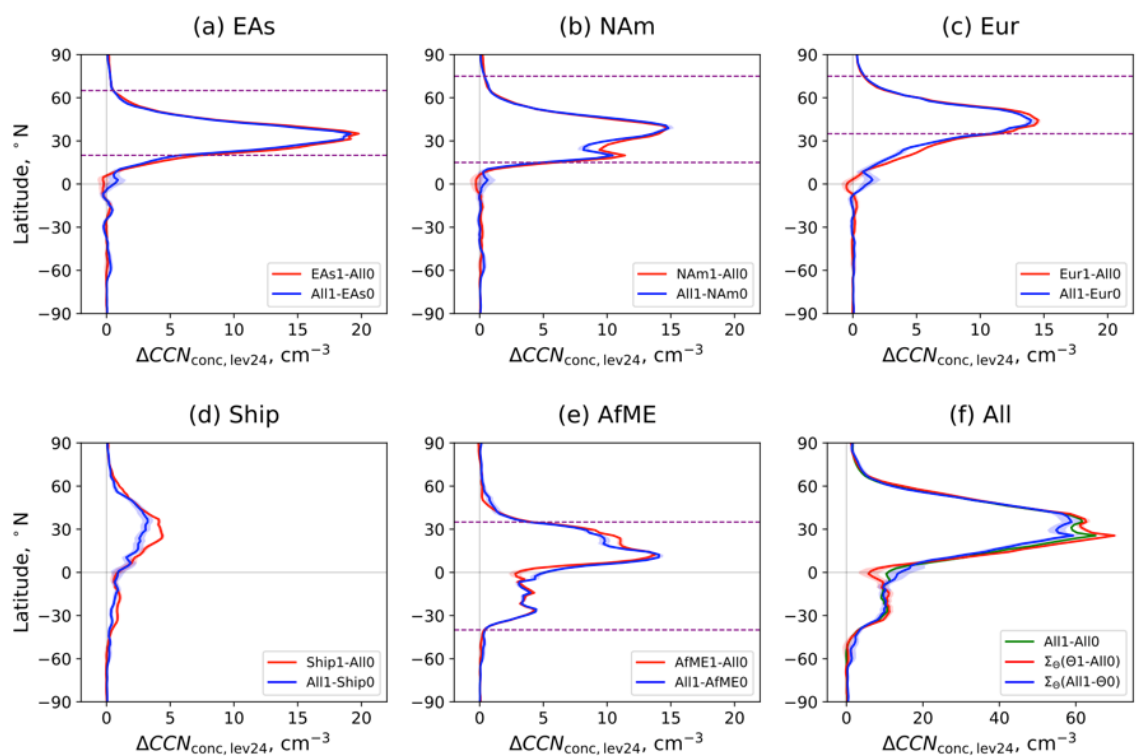

**Figure S9.** Differences in zonal-mean annual-mean global-mean CCN concentration at a supersaturation of 0.1% in model level 24 (~860hPa, in the lower troposphere) for different combinations of scenarios. See the Fig. 7 caption for an explanation of the figure components.

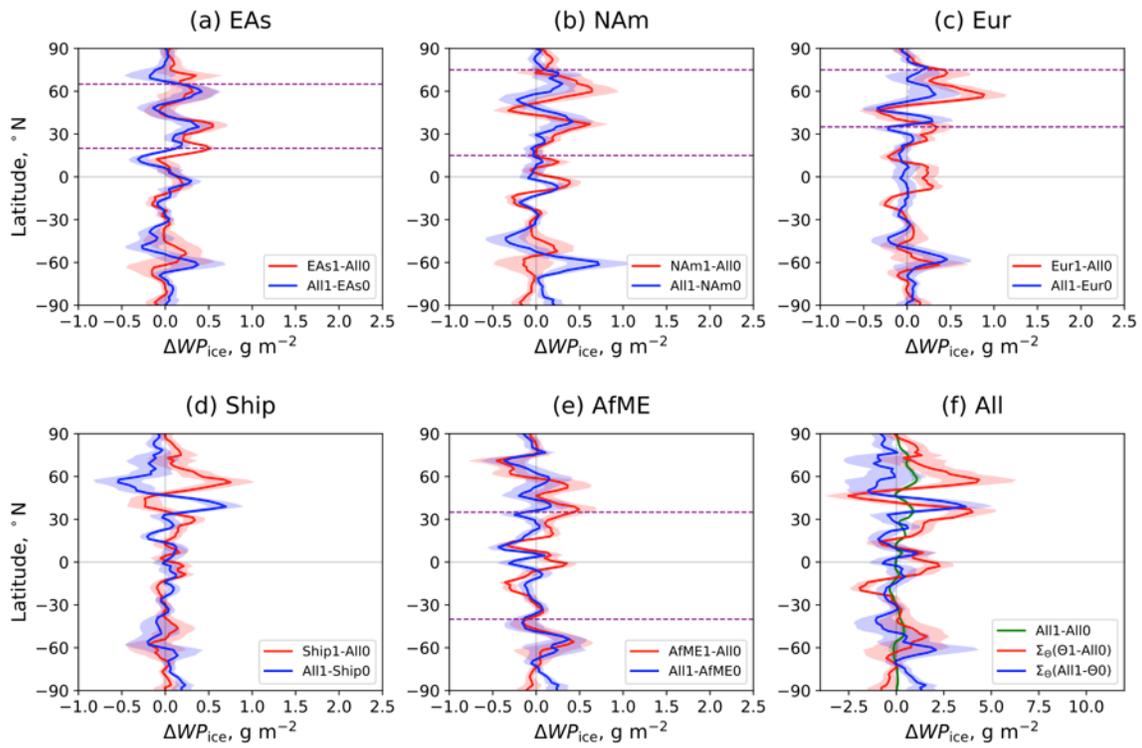

**Figure S10.** Differences in zonal-mean annual-mean grid-box ice water path for different combinations of scenarios. See the Fig. 7 caption for an explanation of the figure components.

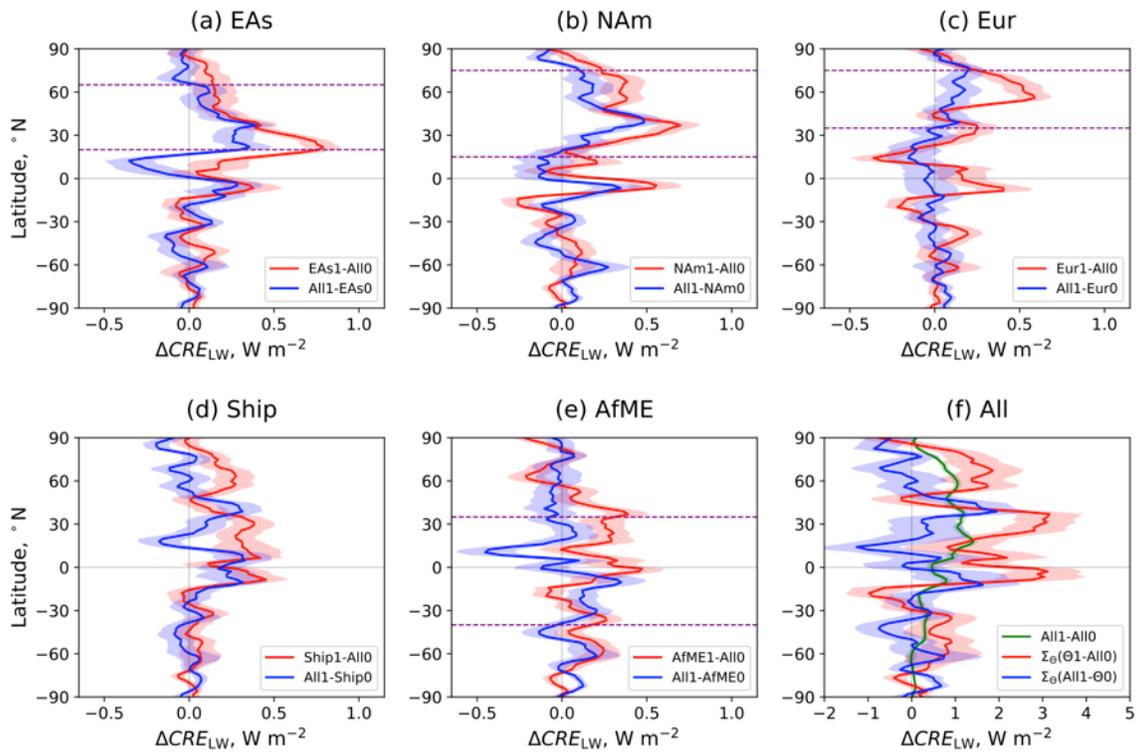

**Figure S11.** Differences in zonal-mean annual-mean longwave cloud radiative effect for different combinations of scenarios. See the Fig. 7 caption for an explanation of the figure components.

Global primary sulfate emissions for scenario All1

| | Intended sources | Actual sources used | Intended emissions, Tg(S) yr$^{-1}$ | Actual emissions used, Tg(S) yr$^{-1}$ | Error (actual - intended), Tg(S) yr$^{-1}$ |
|---|---|---|---|---|---|
| Accumulation mode, surface | shipping, waste treatment, agricultural waste burning | shipping, waste treatment, agricultural waste burning | 0.142 | 0.142 | 0.000 |
| Accumulation mode, elevated | energy, industry, wildfire | energy, industry, wildfire | 1.204 | 1.204 | 0.000 |
| Aitken mode, surface | domestic, transport | shipping, waste treatment, agricultural waste burning | 0.157 | 0.142 | -0.015 |
| Aitken mode, elevated | continuous volcanic | continuous volcanic | 0.157 | 0.157 | 0.000 |

**Figure S12.** Global emissions of primary sulfate aerosol for scenario *All1*. Yellow highlighting indicates the bug in the primary sulfate aerosol emissions (Text S1).

Global primary sulfate and total sulfur emissions for the different scenarios

| | Actual emissions used (global total) | | Error (actual - intended; global total) | | | Error over land, ocean, and coastal regions | | |
|---|---|---|---|---|---|---|---|---|
| | Sulfate, Tg(S) yr$^{-1}$ | Total sulfur, Tg(S) yr$^{-1}$ | Error, Tg(S) yr$^{-1}$ | Relative to intended sulfate emissions | Relative to intended sulfur emissions | Land-only error, Tg(S) yr$^{-1}$ | Ocean-only error, Tg(S) yr$^{-1}$ | Coastal error, Tg(S) yr$^{-1}$ |
| All1  | 1.645 | 84.596 | -0.015 | -0.93%  | -0.02% | -0.093 | +0.101 | -0.024 |
| All0  | 0.363 | 32.718 | +0.000 | +0.00%  | +0.00% | +0.000 | +0.000 | +0.000 |
| Ship1 | 0.640 | 38.395 | +0.138 | +27.62% | +0.36% | +0.002 | +0.103 | +0.034 |
| EAs1  | 0.584 | 43.435 | -0.048 | -7.62%  | -0.11% | -0.031 | -0.000 | -0.017 |
| NAm1  | 0.601 | 42.941 | -0.018 | -2.96%  | -0.04% | -0.010 | -0.000 | -0.008 |
| Eur1  | 0.516 | 39.670 | -0.022 | -4.04%  | -0.05% | -0.013 | -0.000 | -0.009 |
| AfME1 | 0.481 | 38.054 | -0.016 | -3.22%  | -0.04% | -0.009 | -0.000 | -0.006 |
| CAs1  | 0.455 | 37.070 | -0.017 | -3.54%  | -0.05% | -0.015 | -0.000 | -0.002 |
| SAs1  | 0.430 | 35.949 | -0.014 | -3.09%  | -0.04% | -0.010 | -0.000 | -0.003 |
| SEAs1 | 0.408 | 34.815 | -0.007 | -1.74%  | -0.02% | -0.002 | -0.000 | -0.005 |
| SAm1  | 0.403 | 34.755 | -0.011 | -2.70%  | -0.03% | -0.005 | -0.000 | -0.006 |
| ANZ1  | 0.393 | 33.965 | -0.001 | -0.20%  | -0.00% | -0.000 | -0.000 | -0.001 |
| Ship0 | 1.368 | 78.919 | -0.154 | -10.11% | -0.19% | -0.095 | -0.001 | -0.058 |
| EAs0  | 1.424 | 73.879 | +0.033 | +2.36%  | +0.04% | -0.063 | +0.102 | -0.006 |
| NAm0  | 1.407 | 74.373 | +0.003 | +0.21%  | +0.00% | -0.083 | +0.102 | -0.016 |
| Eur0  | 1.492 | 77.645 | +0.006 | +0.42%  | +0.01% | -0.080 | +0.102 | -0.015 |
| AfME0 | 1.527 | 79.260 | +0.001 | +0.04%  | +0.00% | -0.084 | +0.102 | -0.017 |
| CAs0  | 1.552 | 80.245 | +0.001 | +0.08%  | +0.00% | -0.079 | +0.101 | -0.021 |
| SAs0  | 1.577 | 81.366 | -0.002 | -0.11%  | -0.00% | -0.083 | +0.101 | -0.020 |
| SEAs0 | 1.599 | 82.500 | -0.008 | -0.51%  | -0.01% | -0.091 | +0.102 | -0.018 |
| SAm0  | 1.605 | 82.559 | -0.004 | -0.26%  | -0.01% | -0.088 | +0.102 | -0.018 |
| ANZ0  | 1.614 | 83.349 | -0.015 | -0.90%  | -0.02% | -0.093 | +0.101 | -0.023 |

**Figure S13.** Global emissions of primary sulfate aerosol and total sulfur (sulfur dioxide, primary sulfate, DMS) for the different scenarios. The errors between the "actual" emissions and the "intended" emissions are due to the bug in the primary sulfate aerosol emissions (Text S1). Orange bars illustrate the relative magnitudes of the values in each cell, scaled by the largest value in the column. Blue and red bars illustrate the relative magnitudes of the errors.

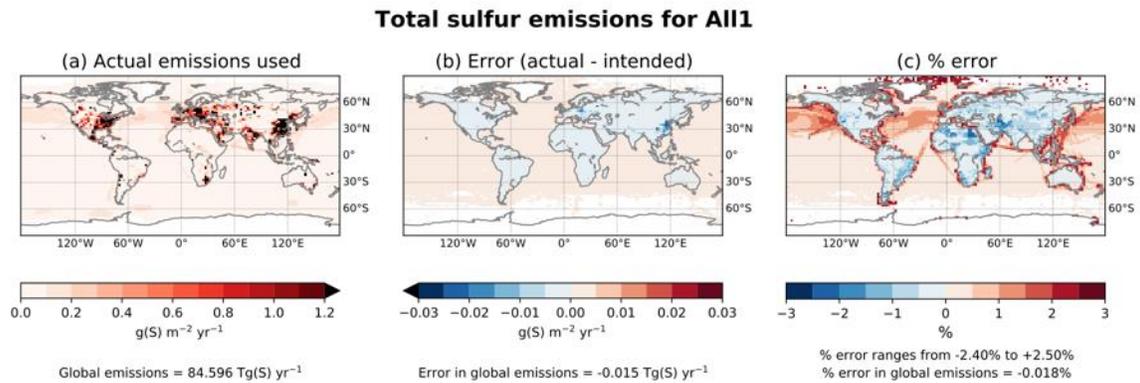

**Figure S14.** Emissions of total sulfur (sulfur dioxide, primary sulfate, DMS) for scenario *All1*. The errors between the "actual" emissions and the "intended" emissions are due to the bug in the primary sulfate aerosol emissions (Text S1)

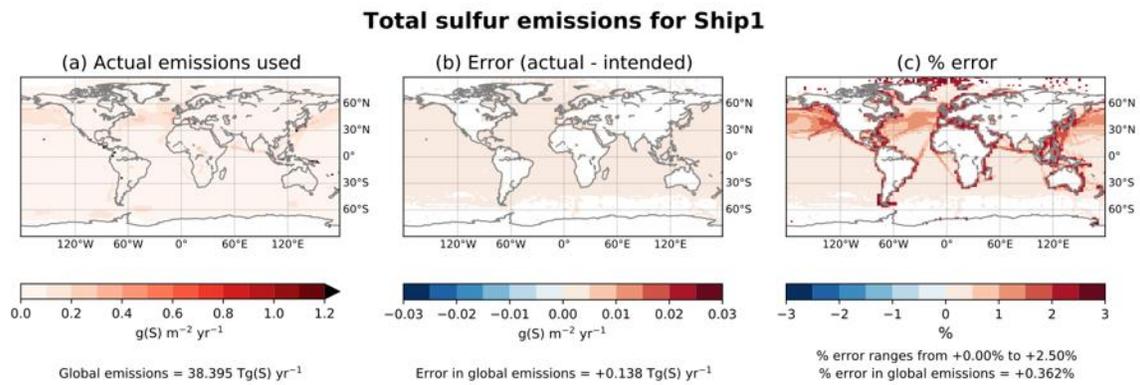

**Figure S15.** Emissions of total sulfur for scenario *Ship1*. See the Fig. S14 caption for further details.

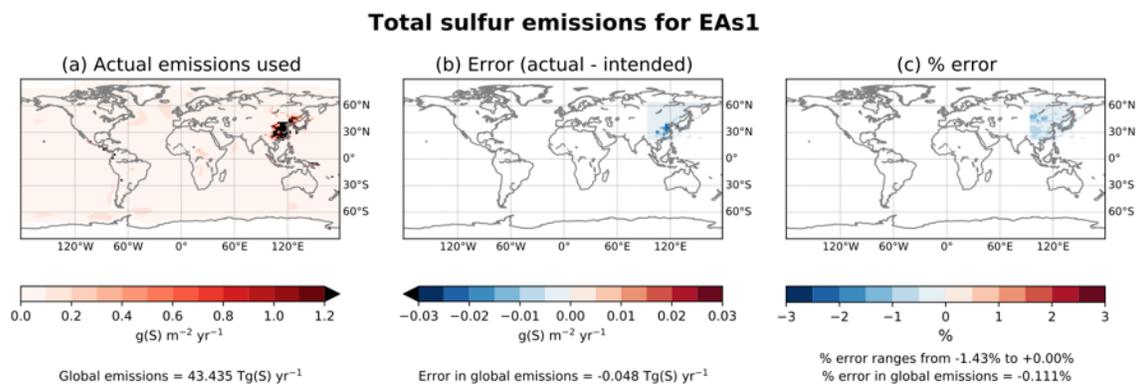

**Figure S16.** Emissions of total sulfur for scenario *EAs1*. See the Fig. S14 caption for further details.

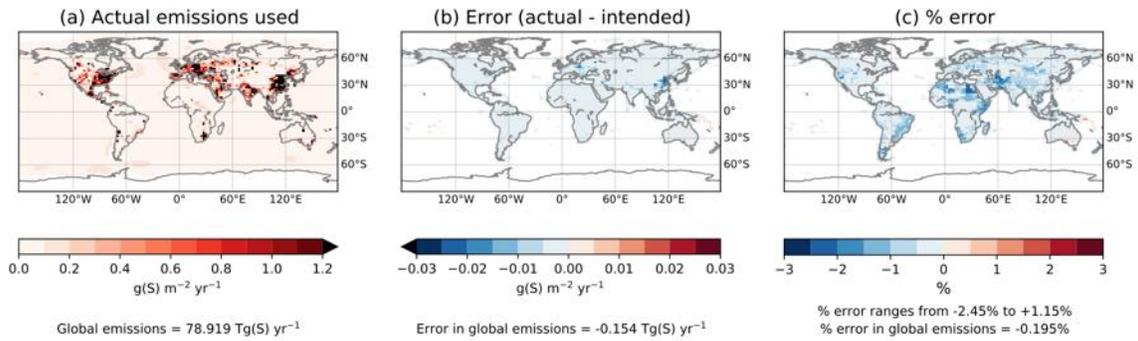

**Figure S17.** Emissions of total sulfur for scenario *Ship0*. See the Fig. S14 caption for further details.

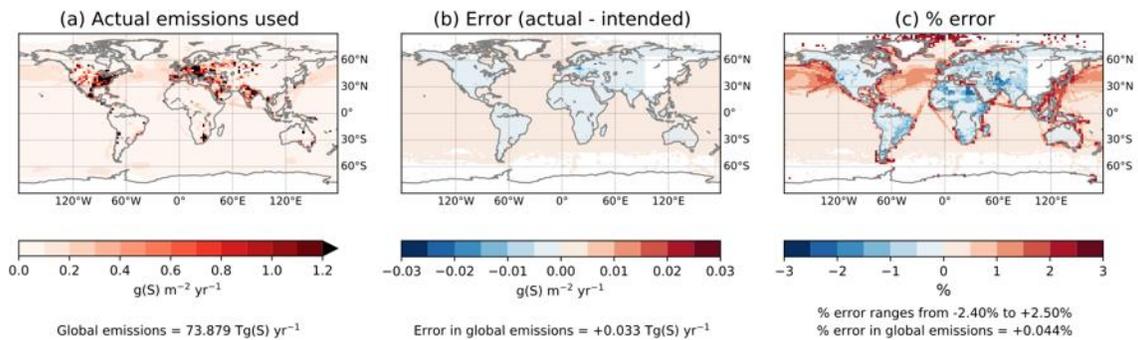

**Figure S18.** Emissions of total sulfur for scenario *Eas0*. See the Fig. S14 caption for further details.

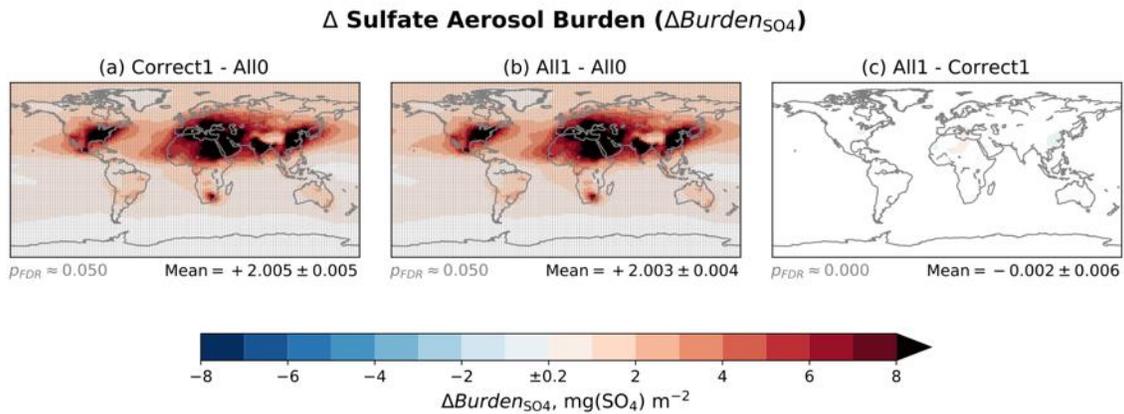

**Figure S19.** Differences in annual-mean sulfate aerosol burden for (a) *Correct1–All0*, (b) *All1–All0*, and (c) *All1–Correct1*. The *All1–Correct1* difference indicates the impact of the error between the "actual" emissions and the "intended" emissions, due to the bug in the primary sulfate aerosol emissions (Text S1). White indicates differences with a magnitude smaller than the value in the centre of the colour bar (±0.2). Stippling indicates statistical significance, calculated using a two-sample *t*-test, using a significance level of 0.05 with the false discovery rate being controlled (Benjamini & Hochberg, 1995; Wilks, 2016); the approximate *p*-value threshold ($p_{FDR}$) is written underneath each map. Area-weighted mean differences are written underneath each map.

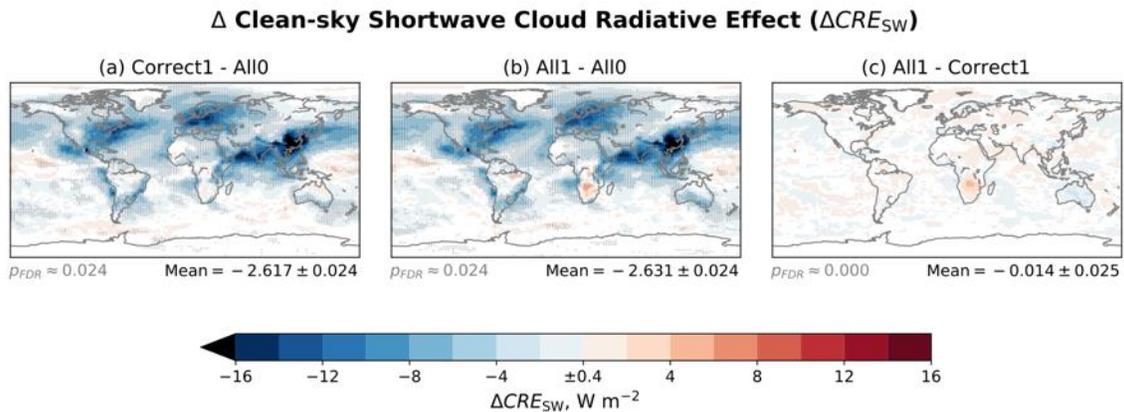

**Figure S20.** Differences in annual-mean clean-sky shortwave cloud radiative effect for (a) *Correct1–All0*, (b) *All1–All0*, and (c) *All1–Correct1*. See the Fig. S19 caption for an explanation of the figure components.

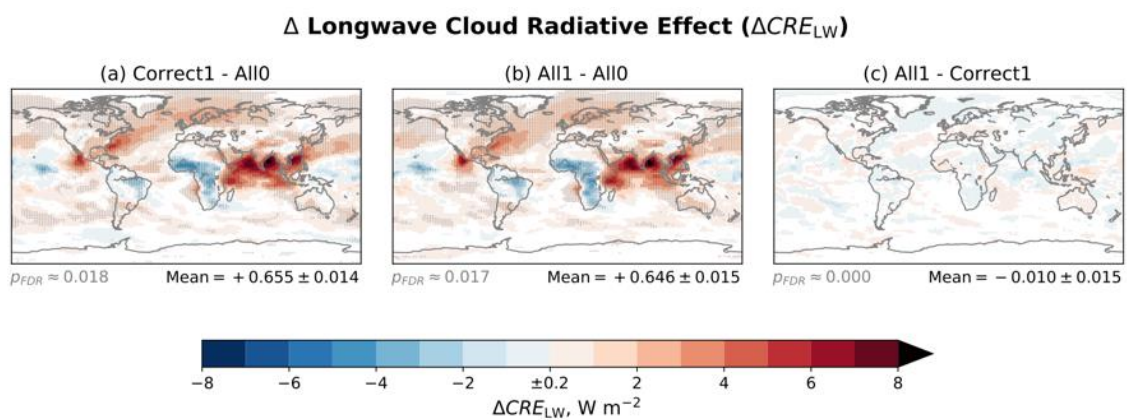

**Figure S21.** Differences in annual-mean longwave cloud radiative effect for (a) *Correct1–All0*, (b) *All1–All0*, and (c) *All1–Correct1*. See the Fig. S19 caption for an explanation of the figure components.